\documentclass[aps,prd,preprintnumbers,groupedaddress,nofootinbib,amssymb,eqsecnum,notitlepage]{revtex4-1}

\usepackage{here}
\usepackage[utf8]{inputenc}
\usepackage{color}
\usepackage{amsfonts}
\usepackage{graphicx}
\usepackage{dsfont} 

\usepackage{color}
\usepackage{bm}
\usepackage{graphicx}
\usepackage{amsmath}
\usepackage{amssymb}
\usepackage{enumitem}
\usepackage{hyperref}
\usepackage{subfigure}
\usepackage{array}
\usepackage[english]{babel}
\usepackage{tensor}
\usepackage{comment}
\usepackage[normalem]{ulem}
\usepackage[dvipsnames]{xcolor}
\usepackage{hyperref}
\usepackage{comment}

\allowdisplaybreaks

\def\be{\begin{equation}}
\def\ee{\end{equation}}
\def\ba{\begin{eqnarray}}
\def\ea{\end{eqnarray}}

\def\f{\frac}

\newcommand{\lsim}{\mathrel{\hbox{\rlap{\lower.55ex \hbox{$\sim$}} \kern-.3em \raise.4ex \hbox{$<$}}}}
\usepackage{tikz}

\newcommand\lcdm{$\Lambda$CDM}
\newcommand{\planck}{\textsl{Planck}}

\def\eftcamb{\texttt{EFTCAMB}}
\def\camb{\texttt{CAMB}}
\def\eftcosmomc{\texttt{EFTCosmoMC}}

\begin{document}

\preprint{YITP-21-138, WUCG-21-14}

\title{Probing Modified Gravity with Integrated Sachs-Wolfe 
CMB and \\
Galaxy Cross-correlations}

\smallskip

\author{Joshua A. Kable$^{1}$, Giampaolo Benevento$^{1}$, Noemi Frusciante$^{2}$, Antonio De Felice$^{3}$, 
Shinji Tsujikawa$^{4}$}

\smallskip

\affiliation{
$^{1}$Johns Hopkins University 3400 North Charles Street Baltimore, MD 21218, USA
\smallskip\\
$^{2}$Instituto de Astrof\'isica e Ci\^encias do Espa\c{c}o, Faculdade de Ci\^encias da Universidade de Lisboa,  \\ Edificio C8, Campo Grande, P-1749016, Lisboa, Portugal 
\smallskip \\
$^{3}$Center for Gravitational Physics, Yukawa Institute for Theoretical Physics, Kyoto University, 606-8502, Kyoto, Japan
\smallskip\\
$^4$Department of Physics, Waseda University, 3-4-1 Okubo, Shinjuku, Tokyo 169-8555, Japan
}
 
\smallskip

\begin{abstract}
We use the cross-correlation power spectrum of the integrated Sachs-Wolfe (ISW) effect in the cosmic microwave background (CMB) temperature anisotropy and galaxy fluctuations to probe the physics of late-time cosmic acceleration. For this purpose, we focus on three models of dark energy that belong to a sub-class of Horndeski theories with the speed of gravity equivalent to that of light: Galileon Ghost Condensate (GGC), Generalized Cubic Covariant Galileon (GCCG), and K-mouflage. In the GGC and GCCG models, the existence of cubic-order scalar self-interactions allows a possibility for realizing negative ISW-galaxy cross-correlations, while the K-mouflage model predicts a positive correlation similar to the $\Lambda$-cold-dark-matter ($\Lambda$CDM) model. In our analysis, we fix the parameters of each model to their best-fit values derived from a baseline likelihood analysis with observational data from CMB, baryon acoustic oscillations, and supernovae type Ia. Then we fit those best-fit models to the ISW-galaxy cross-correlation power spectrum extracted from a collection of photometric redshift surveys. We find that both GGC and GCCG best-fit models degrade the fit to the ISW-galaxy cross-correlation data compared to $\Lambda$CDM best-fit model. This is attributed to the fact that, for their best-fit values constrained from the baseline likelihood, the cubic-order scalar self-interaction gives rise to suppressed ISW tails relative to $\Lambda$CDM. The K-mouflage best-fit model is largely degenerate with the $\Lambda$CDM best-fit model and has a positively correlated ISW-galaxy power close to that of $\Lambda$CDM.
\end{abstract}

\date{\today}

\maketitle

\tableofcontents
\section{Introduction}

The observations of supernovae 
type Ia (SNIa) \cite{SupernovaSearchTeam:1998fmf,SupernovaCosmologyProject:1998vns}, along with the measurements of the cosmic microwave background (CMB) \cite{WMAP:2003elm,Planck:2013pxb}, and the
baryon acoustic oscillations (BAO) \cite{SDSS:2005xqv}, have shown that 
the universe is currently undergoing an epoch of cosmic acceleration. 
The unknown source for this phenomenon is dubbed dark energy, which 
has an effective negative pressure $P$ against gravity. 
At the background level dark energy is quantified by an equation of state parameter $w_{\rm DE}=P/\rho$, where $\rho$ is 
a dark energy density.
The simplest model of dark energy is the cosmological constant 
given by the constant density 
$\rho=\Lambda$ \cite{Weinberg:1988cp,Sahni/Starobinsky:2000,Peebles:2002gy}, in which case the equation of state parameter is $w_{\rm DE} = -1$. 
The standard model of cosmology, $\Lambda$CDM, is built on the framework 
of General Relativity (GR) and assumes a homogeneous and isotropic background 
with a cosmological constant, $\Lambda$, and a non-relativistic 
cold dark matter (CDM) fluid. 
With these simple assumptions, the $\Lambda$CDM model has been overall 
successful at explaining observational data over a vast array of 
cosmological epochs (see e.g., \cite{Review_Particle_Physics_short/2020, cyburt/etal:2016,Scolnic_short/etal:2018,Alam_short/etal:2021}).

Despite this success, if the vacuum energy appearing in particle physics is responsible for the cosmological constant, the typical vacuum 
energy scale is enormously higher than the observed dark energy scale
(see e.g., \cite{Weinberg:1988cp,Martin:2012bt,Joyce:2014kja,Padilla:2015aaa} for reviews). Moreover, there are currently tensions on the values 
of some cosmological parameters between different data sets. 
The tension on the value of today's cosmic expansion rate
$H_0=100\,h$\,km\,sec$^{-1}$\,Mpc$^{-1}$ in the $\Lambda$CDM model 
is estimated to be 
$\sim 4\sigma$ between the Planck CMB 
observations and its direct measurements at low 
redshifts \cite{Aghanim:2018eyx,Riess/etal:2021} 
(see Refs.~\cite{knox&millea/2020,Riess/2020} for reviews).
Similarly, there is a 2$\sim$3$\sigma$ tension for the determination of the amount of galaxy clustering quantified by the parameter 
$S_8 \equiv \sigma_{8,0} \sqrt{\Omega_{{\rm m},0}/0.3}$, 
where $\sigma_{8,0}$ is present-day amplitude of matter fluctuations 
on scale 8$h^{-1}$ Mpc and 
$\Omega_{{\rm m},0}$ is today's density parameter of nonrelativistic 
matter \cite{Hildebrandt_short/etal:2020,Joudaki_short/etal:2018,Hikage_short/etal:2019,Abbott_short/etal:2021,Heymans_short/etal:2021}. 

There have been many dark energy models proposed so far designed 
to improve the aforementioned problems in 
$\Lambda$CDM (see e.g., \cite{bull_short/etal:2016,Valentino/etal:2021,Schoneberg/etal:2021} for reviews). 
One simple extension is to consider time variations in the dark energy equation of state. 
For instance, the quintessence scenario based on 
a scalar field slowly rolling on a nearly flat 
potential gives rise to time variation of $w_{\rm DE}$ 
in the region 
$w_{\rm DE}>-1$ \cite{Copeland:2006wr,Martin:2008qp,Tsujikawa:2013}.
However, there has been no strong statistical evidence that quintessence models are observationally favored 
over $\Lambda$CDM \cite{Chiba:2012cb,Durrive:2018quo}. Moreover, it has been shown that the quintessence parameter space $w_{\rm DE}>-1$ is expected to prefer a lower value of $H_0$ meaning it would worsen the Hubble tension (See e.g. Ref. \cite{Banerjee/etal:2021}). 

Alternatively, one can consider cosmological models that exhibit deviations from GR on scales 
relevant to the present-day cosmic acceleration.
From Lovelock's theorem~\cite{Lovelock:1971yv,Lovelock:1972vz}, 
any infrared departure from GR should include 
new degrees of freedom (DOFs).  
The new DOFs can be a scalar field, a vector 
field, or a massive graviton.
Among them, the scalar field is the simplest example 
that is compatible with the cosmological dynamics of the homogeneous and isotropic background. 
In so-called scalar-tensor theories where the scalar 
field is coupled to gravity through nonminimal and 
derivative couplings, there have been numerous attempts
for constructing viable modified gravity (MG) models of late-time cosmic 
acceleration \cite{Boisseau:2000pr,Lue:2004rj,Copeland:2006wr,Silvestri:2009hh,Nojiri:2010wj,Tsujikawa:2010zza,Capozziello:2011et,Clifton:2011jh,Bamba:2012cp,Joyce:2014kja,Koyama:2015vza,Avelino:2016lpj,Joyce:2016vqv,Nojiri:2017ncd,Kase:2018iwp,Ferreira:2019xrr,Kobayashi:2019hrl,Frusciante:2019xia,CANTATA:2021ktz}. 
Unlike quintessence, it is possible to realize the 
phantom dark energy equation of state ($w_{\rm DE}<-1$) 
without having ghosts.
Many of such models belong to a framework of Horndeski
theories \cite{Horndeski:1974wa} in which the Lagrangian is constructed to 
include the gravitational and scalar-field equations 
of motion up to second-order 
derivatives (see also 
Refs.~\cite{Deffayet:2011gz,Kobayashi:2011nu,Charmousis:2011bf}).

After the gravitational-wave event 
GW170817 \cite{LIGOScientific:2017vwq}, the speed of gravitational 
waves $c_t$ is constrained to be very close to that of light $c$.
The Lagrangian of Horndeski theories 
realizing $c_t=c$ is restricted to be 
of the form $L=G_2(\phi,X)+G_3(\phi, X)\square \phi+G_4(\phi)R$ \cite{Kobayashi:2011nu,DeFelice:2011bh}, 
where $R$ is the Ricci scalar, $G_4$ is a function of the scalar field $\phi$, and $G_2$ and $G_3$ depend on 
$\phi$ as well as $X=\nabla^{\mu}\phi \nabla_{\mu}\phi$ 
(with $\nabla^{\mu}$ being the covariant derivative operator).
Even in this subclass of Horndeski theories, there are 
several models of dark energy showing a better fit to 
the SNIa, CMB, and BAO data in comparison to $\Lambda$CDM.

One of such examples is the Galileon Ghost Condensate (GGC) 
model \cite{Deffayet:2010qz,Kase:2018iwp} characterized by the functions 
$G_2(X)=a_1X+a_2 X^2$, $G_3(X)=3a_3 X$, and $G_4(X)=M_{\rm pl}^2/2$, 
where $M_{\rm pl}$ is the reduced Planck mass.
This includes the Cubic Galileon (G3) \cite{Deffayet:2009wt}  
as a specific case ($a_2=0$), 
which is disfavored by
data \cite{Nesseris:2010pc,Renk:2017rzu,Peirone:2017vcq} 
mostly due to the large deviation of 
$w_{\rm DE}$ from $-1$ for a tracker solution 
($w_{\rm DE}=-2$ during the matter era).
The existence of the term $a_2 X^2$ in the GGC model allows 
a phantom dark energy equation of state between $-2$ and $-1$.
The large-scale CMB temperature anisotropy can be suppressed 
by the cubic interaction term $3a_3 X \square \phi$ relative 
to $\Lambda$CDM. These properties lead to a statistical preference 
of GGC over $\Lambda$CDM, when the SNIa, CMB, and BAO data are used in the analysis \cite{Peirone:2019aua}.
A similar statistical preference \cite{Frusciante:2019puu} 
is also present in the Generalized Cubic Covariant Galileon (GCCG) 
model \cite{DeFelice:2011bh,DeFelice:2011aa} characterized by the functions 
$G_2(X)=-c_2 M_2^{4(1-p)} (-X/2)^{p}$, 
$G_3(X)=-c_3 M_3^{1-4p_3} (-X/2)^{p_3}$, 
and $G_4=M_{\rm pl}^2/2$, which also realizes 
the phantom dark energy equation of state.

There are also dark energy models in which the scalar field has 
a nonminimal coupling $G_4(\phi)R$ to the Ricci scalar. 
One of such examples is 
K-mouflage \cite{Brax:2014wla,Brax:2015dma,Brax:2016Km}, 
where the corresponding 
Lagrangian in the Jordan frame is of the form 
$L=G_2(\phi,X)+G_4(\phi)R$.
The nonminimal coupling gives rise to the propagation of 
an effective fifth force
between the scalar field and baryons \cite{Joyce:2014kja}, 
so it is necessary to implement a screening mechanism 
to hide the extra force in local regions of the universe.
Unlike the chameleon mechanism \cite{Khoury:2003aq} or 
Vainshtein mechanism \cite{Vainshtein:1972sx} in which a scalar potential 
$V(\phi)$ or a scalar self-interaction $G_3(X) \square \phi$ 
suppresses the effective fifth force in the region of 
high densities, K-mouflage resorts to 
nonlinear kinetic terms in $G_2(X)$ for the screening mechanism to
work. On today's Hubble scales, such nonlinear kinetic terms can drive 
cosmic acceleration as in the K-essence scenario \cite{Chiba:1999ka,Armendariz-Picon:2000nqq}.

On scales relevant to the linear growth of large-scale structures, 
the presence of Lagrangians $G_3(X) \square \phi$ and $G_4(\phi)R$  
gives rise to an effective gravitational coupling $G_{\rm eff}$
with matter different from the Newton gravitational constant 
$G_{\rm N}$ \cite{DeFelice:2011hq,Kase:2018aps} 
(see also Refs.~\cite{Amendola:2007rr,Bean:2010zq,Silvestri:2013ne,Amendola:2019laa}). Then, the growth of matter density contrast is 
modified from that in $\Lambda$CDM, whose property can be used to 
place constraints on MG models from the growth rate measurements 
like redshift-space distortions (RSDs).

In MG models, the two gravitational potentials $\Psi$ and 
$\Phi$, which appear as the temporal and spatial components of metric 
perturbations in the longitudinal gauge, are not generally 
equivalent to each other.
Thus, the relation between the lensing potential 
$\Phi_{\rm eff}=\Psi+\Phi$ and the density contrast 
$\delta_{\rm m}$ is subject to modification in comparison 
to $\Lambda$CDM \cite{Amendola:2007rr,Bean:2010zq,Silvestri:2013ne,Amendola:2019laa}. 
As such, the MG models not only affect the weak lensing potential  \cite{Acquaviva:2005xz,Carbone:2013dna} but also modify the 
low-multipole CMB temperature anisotropy through 
the integrated Sachs-Wolfe (ISW) effect \cite{Sachs:1967er,Kofman:1985fp}. 
As a combined effect, the cross-correlation between the ISW effect in CMB 
and large-scale structure (LSS) \cite{Giannantonio:2008zi} is 
affected as well.
The ISW signal appears in the CMB power spectrum at large angular scales corresponding to multipoles $l \lesssim 40$, where the primary Sachs-Wolfe effect and cosmic variance are the dominant sources of anisotropy. This prevents the detection of the ISW signal using CMB data alone.
However, such a signal can be isolated by cross-correlating the CMB data with LSS tracers \cite{Boughn:1997vs,Boughn/etal:2001,Stolzner:2017ged}. 
This provides a powerful tool to constrain
MG models 
of dark energy \cite{Song/etal:2007,Barreira/etal:2012,Renk:2017rzu,Frusciante:2019puu,Giacomello:2018jfi,Hang/etal:2021b}.

In this paper, we are interested in constraining the ISW effect due to 
modified gravitational interactions through the cross-correlation between CMB and galaxy surveys. To construct an ISW-galaxy cross-correlation likelihood, we extend to MG models the tomographic analysis of the ISW signal using photometric measurements of the redshift of galaxies applied to the \lcdm\ model (see e.g., \cite{Scranton/etal:2003,Sawangwit/etal:2010,Francis_and_Peacock:2010,Stolzner:2017ged,Hang/etal2021,Krolewski/etal:2021}). 
We specialize our analysis to three the MG models mentioned above, 
i.e., GGC, GCCG, and K-mouflage.
We show that the best-fit parameters in the GGC and GCCG models 
constrained by the SNIa, CMB, and BAO data give rise to a suppressed 
ISW power induced by the cubic interaction $G_3(X) \square \phi$. 
This leads to worse fits to the ISW-galaxy cross-correlation 
data obtained from a collection of photometric redshift surveys that were originally combined in Ref.~\cite{Stolzner:2017ged} for an ISW analysis assuming $\Lambda$CDM. 
Hence the ISW-galaxy cross-correlation power spectrum provides strong constraints 
on dark energy models containing the cubic derivative interaction
$G_3(X)\square \phi$. The best-fit parameters for K-mouflage are found to be  largely degenerate with the \lcdm\ case, meaning that 
we essentially recover the \lcdm\ limit. 

This paper is organized as follows. 
In Sec.~\ref{Sec:theory}, we provide a review of the cross-correlation 
power spectrum between the ISW effect in CMB and galaxy clusterings.
In Sec.~\ref{Sec:models},  we briefly revisit the dynamics of the background and perturbations in GGC, GCCG, and K-mouflage models.
In Sec.~\ref{Sec:method}, we outline the methodology adopted to probe 
our MG models with the ISW-galaxy cross-correlation power spectrum.  
In Sec.~\ref{Sec:Results}, we present and discuss the results of our analysis for each model. Finally, in Sec.~\ref{Sec:Conclusion}, we offer conclusions. 
Throughout the paper, we use the natural units where the speed of 
light $c$ and the reduced Planck constant $\hbar$ 
are equivalent to 1.

\section{ISW-galaxy cross-correlations}\label{Sec:theory}

We first review the power spectrum of ISW-galaxy 
cross-correlations in MG theories. 
This general prescription accommodates not only dark energy 
models in GR \cite{Giannantonio:2008zi}, but also those in scalar-tensor \cite{Kimura:2011td} 
and vector-tensor theories \cite{Nakamura:2018oyy}.
Let us consider the perturbed line element on the spatially flat 
Friedmann-Lema{\^i}tre-Robertson-Walker (FLRW) background 
in the Newtonian gauge:
\be
{\rm d}s^2=-\left( 1+2\Psi \right) {\rm d}t^2
+a^2(t) \left(1-2\Phi \right) \delta_{ij} {\rm d}x^i 
{\rm d}x^j\,,
\label{permet}
\ee
where $\Psi$ and $\Phi$ are the gravitational potentials 
that depend on time $t$ and spatial position $x^i$, 
and $a(t)$ is the time-dependent scale factor. 

The effective gravitational potentials associated with 
the bending of light rays is given by 
\be
\Phi_{\rm eff}=\Psi+\Phi\,.
\ee
The ISW effect of CMB temperature anisotropy
occurs by the time variation of $\Phi_{\rm eff}$ after the recombination epoch.
The ISW contribution $\delta T_{\rm ISW}$ to the CMB temperature perturbation
divided by its average temperature $\bar{T}$ can be quantified as
\be
\f{\delta T_{\rm ISW}(\hat{n})}{\bar{T}}=
-\int^{z_*}_{0}{\rm d}z\,e^{-\tau(z)}
\f{\partial \Phi_{\rm eff}}{\partial z}(z,\hat{n}\chi (z))\,,
\label{Tper}
\ee
where $\hat{n}$ is a unit vector along the line of sight, 
and $z=1/a-1$ is the redshift with the value $z_*$ 
at recombination, and $\tau$ is  
the visibility function.

The fluctuations of the angular distribution of galaxies, 
with the average number $\bar{N}_{\rm g}$, 
are expressed in the form 
\be
\f{\delta N_{\rm g}(\hat{n})}{\bar{N}_{\rm g}}=\int^{z_*}_0 {\rm d}z\,\delta_{\rm g}(z,x^i)W(z)\,,
\label{Nper}
\ee
where $W(z)$ is the selection function of the survey, and $\delta_{\rm g}$ is the galaxy number density contrast. 
We relate $\delta_{\rm g}$ with the matter density contrast $\delta_{\rm m}$, as
\be
\delta_{\rm g}(z,x^i)=b\, 
\delta_{\rm m}(z,\hat{n}\chi (z))\,,
\ee
where $b$ is a bias factor, and 
$\chi(z)=\int_0^z {\rm d}\tilde{z}/H(\tilde{z})$ 
is a comoving distance. 
The selection function $W(z)$, which satisfies 
the normalization $\int_0^\infty {\rm d}z\,W(z)=1$, is specific of the galaxy survey considered. The typical choice of $W(z)$ approximating observed galaxy distributions 
is given by 
\be
W(z) = \frac{\beta}{\Gamma[(\alpha+1)/\beta]} \left( \frac{z}{z_{0}} \right)^{\alpha} 
\exp\left[ -\left( \frac{z}{z_{0}} \right)^{\beta} \right]\,,
\label{window}
\ee
where $\Gamma[x]$ is the gamma function and 
$\alpha, \beta, z_{0}$ are positive constants.
For instance, the 2MASS and SDSS galaxy catalogues can be fitted by Eqn.~(\ref{window}) with 
the constants
$(z_{0}, \alpha, \beta) = (0.072, 1.901, 1.752)$
and 
$(z_{0}, \alpha, \beta) = (0.113, 3.457, 1.197)$, 
respectively \cite{Giannantonio:2008zi}.

We expand the perturbations (\ref{Tper}) and 
(\ref{Nper}) in terms of the spherical harmonics
$Y_{lm} (\hat{n})$, as 
\ba
\f{\delta T_{\rm ISW}(\hat{n})}{\bar{T}}
&=&\int_0^{z_*}{\rm d}z 
\f{\delta T_{\rm ISW}}{\bar{T}}
(z, \chi \hat{n})
=\sum_{l,m}a_{lm}^{\rm ISW}
Y_{lm} (\hat{n})\,,\\
\f{\delta N_{\rm g}(\hat{n})}{\bar{N}_{\rm g}}
&=& \int_0^{z_*}{\rm d}z 
\f{\delta N_{\rm g}}{\bar{N}_{\rm g}}(z, \chi \hat{n})
=\sum_{l,m}a_{lm}^{\rm g}Y_{lm} (\hat{n})\,,
\ea
where 
$a_{lm}^{\rm ISW}=\int 
{\rm d}\Omega\,(\delta T_{\rm ISW}(\hat{n})/\bar{T})Y_{lm}^*(\hat{n})$ 
and 
$a_{lm}^{\rm g}=\int {\rm d}\Omega\,(\delta N_{\rm g}(\hat{n})/\bar{N}_{\rm g})Y_{lm}^*(\hat{n})$,
with $\Omega$ being a solid angle.
We also expand $\delta T_{\rm ISW}(z, \chi \hat{n})/\bar{T}$ and 
$\delta N_{\rm g}(z, \chi \hat{n})/\bar{N}_{\rm g}$ 
in terms of the 
Fourier series, respectively, as
\be
\f{\delta T_{\rm ISW}}{\bar{T}}
(z, \chi \hat{n})=\int \frac{{\rm d}^3k}{(2\pi)^3}
\f{\delta T_{\rm ISW}}{\bar{T}}
(z, {\bm k}) e^{i{\bm k}\cdot \chi \hat{n}}\,,
\qquad
\f{\delta N_{\rm g}}{\bar{N}_{\rm g}}
(z, \chi \hat{n})=\int \frac{{\rm d}^3k}{(2\pi)^3}
\f{\delta N_{\rm g}}{\bar{N}_{\rm g}}
(z, {\bm k}) e^{i{\bm k}\cdot \chi \hat{n}}\,,
\ee
where ${\bm k}$ is a comoving wavenumber, 
with $k=|{\bm k}|$.
Using the property 
$\int {\rm d}\Omega\,e^{i{\bm k}\cdot {\bm r}}Y_{lm}^*(\hat{r})
=4\pi i^l j_l (kr)Y_{lm}^*(\hat{k})$, 
where $j_l (kr)$ is a spherical Bessel function 
with the notations $\hat{r}={\bm r}/r$ and  $\hat{k}={\bm k}/k$,
the coefficients $a_{lm}^{\rm ISW}$ and 
$a_{lm}^{\rm g}$ reduce, respectively, to
\ba
a_{lm}^{\rm ISW} 
&=& 
-\frac{i^l}{2\pi^2} \int_0^{z_*} {\rm d}z_1 
\int {\rm d}^3k_1 e^{-\tau(z_1)}
\f{\partial \Phi_{\rm eff}}{\partial z}(z_1,k_1) 
j_l (k_1 \chi(z_1))Y_{lm}^*(\hat{k}_1)\,,
\label{alm1}\\
a_{lm}^{\rm g} 
&=&
\frac{i^l}{2\pi^2} \int_0^{z_*} {\rm d}z_2 \int {\rm d}^3k_2\,
b\,W(z_2) \delta_{\rm m} (z_2, k_2)j_l (k_2 \chi (z_2))Y_{lm}^*(\hat{k}_2)\,.
\label{alm2}
\ea

The cross-correlation between the ISW and 
galaxy fluctuations is expressed as
\be
\left< \frac{\delta T_{\rm ISW}(\hat{n}_1)}
{\bar{T}} 
\frac{\delta N_{\rm g}(\hat{n}_2)}{\bar{N}_{\rm g}}\right>
= \sum_{l=2}^{\infty} \frac{2 l + 1}{4 \pi} 
C_{l}^{\rm Tg} \mathcal{P}_{l}(\cos\theta) \,,
\ee
where $\mathcal{P}_{l}$ is the Legendre 
polynomial with the angle $\theta$ between the unit vectors 
$\hat{n}_1$ and $\hat{n}_2$, 
and $C_{l}^{\rm Tg}$ is the ISW-galaxy cross-correlation amplitude given by 
\be
C_{l}^{\rm Tg} = \left< a_{lm}^{\rm ISW} 
( a_{lm}^{\rm g})^{*} \right>\,.
\label{Cdef}
\ee
It is also useful to define the galaxy 
auto-correlation function as 
\be
\left< \frac{\delta N_{\rm g}(\hat{n}_1)}
{\bar{N}_{\rm g}} 
\frac{\delta N_{\rm g}(\hat{n}_2)}{\bar{N}_{\rm g}}\right>
= \sum_{l=2}^{\infty} \frac{2 l + 1}{4 \pi} 
C_{l}^{\rm gg} \mathcal{P}_{l}(\cos\theta) \,,
\ee
where 
\be
C_{l}^{\rm gg} = \left< a_{lm}^{\rm g} 
( a_{lm}^{\rm g})^{*} \right>\,.
\label{Cgg}
\ee
We consider the case in which the growth factor 
$D$ of linear matter perturbations 
does not depend on the comoving wavenumber $k$. 
Indeed, this property holds for the MG models 
of dark energy discussed later 
in Sec.~\ref{Sec:models} on scales inside 
the sound horizon.
Then, the matter density contrast is expressed 
in the form 
\be
\delta_{\rm m}(z, \bm{k})=D(z) 
\frac{\delta_{\rm m}(0, \bm{k})}{D_0}\,,
\label{delD}
\ee
where $D_0$ is today's value of $D$. 
Today's matter power spectrum $P_{\rm m}$ 
is defined by 
\be
\left< \delta_{\rm m}(0, \bm{k}_1) 
\delta_{\rm m}^*(0, \bm{k}_2) \right>
= (2 \pi)^{3} \delta_{D}^{(3)}
(\bm{k}_1- \bm{k}_2) 
P_{\rm m}(k_1) \,,
\ee
where $\delta_{D}^{(3)}$ is the three-dimensional 
delta function.
We also introduce the quantity 
$\psi_{\rm ISW}$ characterizing 
the time variation of $\Phi_{\rm eff}$, as 
\be
\frac{\partial \Phi_{\rm eff}}{\partial z}
=-\psi_{\rm ISW}
\frac{\delta_{\rm m}(0, \bm{k})}{D_0}\,.
\label{psiISW}
\ee
Substituting Eqs.~(\ref{alm1}) and (\ref{alm2}) 
into Eqn.~(\ref{Cdef}), 
it follows that 
\be
C_l^{\rm Tg}=\frac{2}{\pi} 
\int {\rm d}k\,k^2 P_{\rm m}(k)
I_{l}^{\rm ISW}(k) I_{l}^{\rm g}(k)
\label{CLTg}
\,,
\ee
where
\ba
I_{l}^{\rm ISW}(k) &=& 
\int_0^{z_*} {\rm d}z_1 e^{-\tau(z_1)}\frac{\psi_{\rm ISW}(z, k)}
{D_0}j_l (k \chi(z_1))\,,\\
I_l^{\rm g}(k) &=& 
\int_0^{z_*} {\rm d}z_2\, b\,W(z_2) \frac{D(z)}{D_0}
j_l (k \chi(z_2))\,.
\label{Ig_def}
\ea
Similarly, substituting Eqs.~(\ref{alm1}) and (\ref{alm2}) 
into Eqn.~(\ref{Cgg}) gives
\be
C_l^{\rm gg}=\frac{2}{\pi} 
\int {\rm d}k\,k^2 P_{\rm m}(k)
I_{l}^{\rm g}(k)^2 
\label{CLgg}
\,.
\ee
The gravitational potentials $\Psi$ and $\Phi$ are 
sourced by the matter density contrast $\delta_{\rm m}$  
through the perturbed Einstein equations. 
In Fourier space without the neutrino's anisotropic stress, 
the Poisson and lensing equations are given, respectively, by \cite{Amendola:2007rr,Bertschinger:2008zb,Pogosian:2010tj}
\ba 
&&-k^2\Psi=4\pi G_{\rm N} a^2\mu(z,k)
\rho_{\rm m}\delta_{\rm m}\,, 
\label{mudef}\\
&&-k^2\Phi_{\rm eff}=8\pi G_{\rm N} 
a^2\Sigma(z,k)\rho_{\rm m}\delta_{\rm m}\,,\label{sigmadef}
\ea
where $G_{\rm N}$ is the Newton's gravitational constant, and 
the dimensionless quantities $\mu \equiv G_{\rm eff}/G_{\rm N}$ 
and $\Sigma$ characterize the effective dimensionless 
gravitational couplings felt by matter and  light, respectively. 
In terms of the gravitational slip parameter 
$\eta=\Phi/\Psi$, the relation between $\mu$ and $\Sigma$ is
\be
\Sigma=\frac{\eta+1}{2}\mu\,.
\ee
The $\Lambda$CDM model gives $\mu=\Sigma=\eta=1$. 
In MG theories, however, $\mu$ and $\Sigma$ 
are generally different from 1. 
This leaves imprints for the growth-rate measurements  
as well as the ISW-galaxy cross-correlation 
power spectrum.

Let us consider the case in which the matter fields are minimally coupled to gravity. Then, the background nonrelativistic matter density evolves as 
\be
\rho_{\rm m}=\rho_{{\rm m},0} (1+z)^3
=\frac{3H_0^2}{8\pi G_{\rm N}}
\Omega_{{\rm m},0} (1+z)^3
\label{rhomdef}
\ee
where $\rho_{{\rm m},0}$ is today's value of $\rho_{\rm m}$, and 
$\Omega_{{\rm m,}0}=8\pi G_{\rm N}\rho_{{\rm m},0}/(3H_0^2)$.
On using Eqs.~(\ref{delD}), (\ref{sigmadef}) and (\ref{rhomdef}), 
it follows that 
\be
\Phi_{\rm eff}=-\frac{3H_0^2 \Omega_{{\rm m},0}}{k^2} (1+z) 
D \Sigma \frac{\delta_{\rm m} (0, {\bm k})}{D_0}\,.
\label{Phieeq}
\ee
Taking the $z$ derivative of Eqn.~(\ref{Phieeq}) and comparing it 
with Eqn.~(\ref{psiISW}), we can express $\psi_{\rm ISW}$ in the form 
\be
\psi_{\rm ISW}=\frac{3H_0^2 \Omega_{{\rm m},0}}{k^2}
D \Sigma {\cal F}\,,
\label{psiISWa}
\ee
where 
\be
{\cal F} \equiv 1+(1+z) \frac{{\rm d}}{{\rm d}z} 
\ln (D\Sigma)\,.
\label{calF}
\ee
Substituting Eqn.~(\ref{psiISWa}) into Eqn.~(\ref{CLTg}), the ISW-galaxy cross-correlation power spectrum yields
\be
C_l^{\rm Tg}=\frac{6H_0^2 \Omega_{{\rm m},0}}{\pi D_0^2} 
\int {\rm d}k\,P_{\rm m}(k)
\int_0^{z_*} {\rm d} z_1\,e^{-\tau(z_1)} D \Sigma {\cal F}
j_l (k \chi_1) 
\int_0^{z_*} {\rm d}z_2\, b W D 
j_l (k \chi_2)\,, 
\label{CLTg2}
\ee
where $\chi_i \equiv \chi(z_i)$ with $i=1,2$.
For large values of $k$, we employ the Limber 
approximation for an arbitrary $k$-dependent
function $f(k)$, i.e., 
\be
\int {\rm d}k\,k^2 f(k) j_l (k\chi_1) j_l (k\chi_2)
\simeq \frac{\pi}{2} \frac{\delta (\chi_1-\chi_2)}{\chi_1^2} 
f\left( \frac{l_{12}}{\chi_1} \right)\,,
\ee
where $l_{12} \equiv l+1/2$. 
Applying this approximation to Eqn.~(\ref{CLTg2}) together with
the relation ${\rm d} z/{\rm d}\chi=H$, 
it follows that 
\be
C_l^{\rm Tg} \simeq \frac{3H_0^2 \Omega_{{\rm m},0}}
{D_0^2 l_{12}^2} \int_0^{z_*}{\rm d}z\, e^{-\tau} H b 
D^2 \Sigma {\cal F} P_{\rm m} \left( 
\frac{l_{12}}{\chi} \right)\,.
\ee
The necessary condition for the negative ISW-galaxy 
cross-correlation ($C_l^{\rm Tg}<0$) to occur is 
quantified by 
\be
{\cal F}<0\,.
\label{cFcon}
\ee
In terms of the e-folding number 
${\cal N}=\ln a=-\ln (1+z)$, 
the condition (\ref{cFcon}) can be expressed as
\be
{\cal F}=1-\frac{D'({\cal N})}{D({\cal N})}
-\frac{\Sigma'({\cal N})}{\Sigma({\cal N})}<0\,.
\ee
Relating the growth rate $D$ with 
the matter density 
parameter $\Omega_{\rm m}$ and the growth index 
$\gamma$ as $D'({\cal N})/D({\cal N})=(\Omega_{\rm m})^\gamma$, 
the term $D'({\cal N})/D({\cal N})$ is smaller than 1 for 
$0<\Omega_{\rm m}<1$ and $0<\gamma<1$. 
Then, for the realization of the negative ISW-galaxy 
cross-correlation, it is at least necessary to 
satisfy the condition 
\be
\Sigma'({\cal N})>0\,.
\label{Sigmad}
\ee
In GR we have $\Sigma=1$ and hence 
$C_l^{\rm Tg}>0$. 
In MG theories, however, there are 
dark energy models in which $\Sigma$ changes in time, 
so it is possible to realize the negative 
ISW-galaxy cross-correlation.
In Sec.~\ref{Sec:models}, we will present such MG models of dark energy.

\section{Dark energy models}
\label{Sec:models}

In this section, we review dark energy models in the 
framework of cubic-order shift-symmetric Horndeski theories as well as 
K-mouflage theories. 
We first present the quantities $\mu$ and $\Sigma$ 
derived under the quasi-static approximation in 
cubic-order Horndeski theories and apply them to 
GGC and GCCG models.
We then proceed to the case of K-mouflage theories 
in which the nonminimal coupling $G_4(\phi)R$ is 
present besides the K-essence Lagrangian $G_2(\phi,X)$.

\subsection{Shift-symmetric Horndeski theories} 

Let us consider cubic-order shift-symmetric Horndeski theories
given by the action 
\be
{\cal S}=\int {\rm d}^4 x \sqrt{-g} \left[ 
\frac{M_{\rm pl}^2}{2}R+G_2(X)+G_3(X) \square \phi 
\right]+{\cal S}_{\rm m} (\psi_{\rm m}, g_{\mu \nu})\,,
\label{actionHo}
\ee
where $g$ is a determinant of the metric tensor 
$g_{\mu \nu}$, 
$M_{\rm pl}=1/\sqrt{8 \pi G_{\rm N}}$ is 
the reduced Planck mass, $R$ is the Ricci scalar, 
and $G_2$, $G_3$ are functions of 
\be
X=\nabla^{\mu} \phi \nabla_{\mu} \phi\,,
\ee
with a scalar field $\phi$.
Since the action (\ref{actionHo}) does not contain 
the $\phi$ dependence in $G_2$, the scalar field can be 
regarded as a massless field. 
This theory also gives the speed of gravity $c_t$ equivalent to that of 
light \cite{Kobayashi:2011nu,DeFelice:2011bh}, so it evades the observational 
bound on the speed of gravitational waves \cite{LIGOScientific:2017vwq}. 

The action ${\cal S}_{\rm m}$ corresponds to that 
of matter fields $\psi_{\rm m}$. 
We will study the case in which 
the matter sector is described by 
perfect fluids minimally coupled 
to gravity. The background nonrelativistic (pressureless)
matter density $\rho_{\rm m}$ on the FLRW background obeys 
\be
\dot{\rho}_{\rm m}+3H \rho_{\rm m}=0\,,
\ee
where $H=\dot{a}/a$ is the Hubble expansion rate, with a dot 
being the derivative with respect to $t$.
The density $\rho_{\rm r}$ of radiation, which has the 
equation of state $w_{\rm r}=1/3$, satisfies 
the continuity equation $\dot{\rho}_{\rm r}+4H \rho_{\rm r}=0$.
The background gravitational and scalar-field equations of 
motion are given by 
\ba
& &
3M_{\rm pl}^2 H^2=\rho_{\rm DE}+\rho_{\rm m}
+\rho_{\rm r}\,,\label{back1}\\
& &
2 M_{\rm pl}^2 \dot{H}=-\rho_{\rm DE}
-P_{\rm DE}
-\rho_{\rm m}-\frac43 \rho_{\rm r}\,,\label{back2}\\
& &
(G_{2,X}-2 \dot{\phi}^2 G_{2,XX}-6H \dot{\phi} 
G_{3,X}+6H \dot{\phi^3}G_{3,XX}) \ddot{\phi}
+3(H G_{2,X}- \dot{H} \dot{\phi} G_{3,X} 
-3H^2 \dot{\phi} G_{3,X})
\dot{\phi}=0\,,\label{back3}
\ea
where 
\ba
\rho_{\rm DE} &=&
-G_2-2\dot{\phi}^2 G_{2,X}
+6 H \dot{\phi}^3 G_{3,X}\,,\\
P_{\rm DE} &=& 
G_2-2\ddot{\phi} \dot{\phi}^2 G_{3,X}\,.
\ea
The dark energy equation of state is 
defined by 
\be
w_{\rm DE} \equiv \frac{P_{\rm DE}}
{\rho_{\rm DE}}=
-\frac{G_2-2\ddot{\phi} \dot{\phi}^2 G_{3,X}}{G_2+2\dot{\phi}^2 G_{2,X}
-6 H \dot{\phi}^3 G_{3,X}}\,.
\label{wde}
\ee
{}From Eqn.~(\ref{back1}), there is the 
constraint 
\be
\Omega_{\rm DE}+\Omega_{\rm m}
+\Omega_{\rm r}=1\,,
\label{Omecon}
\ee
where 
\be
\Omega_{\rm DE} \equiv 
\frac{\rho_{\rm DE}}{3M_{\rm pl}^2 H^2}\,,
\qquad 
\Omega_{\rm m} \equiv 
\frac{\rho_{\rm m}}{3M_{\rm pl}^2 H^2}\,,
\qquad 
\Omega_{\rm r} \equiv 
\frac{\rho_{\rm r}}{3M_{\rm pl}^2 H^2}\,.
\ee
We also solve Eqs.~(\ref{back2}) and (\ref{back3}) for $\dot{H}$ and $\ddot{\phi}$ to know the dynamics of $H$ and $\phi$.

For the perturbations in the matter sector, we will 
consider the nonrelativistic 
matter perturbation $\delta \rho_{\rm m}$ and velocity potential $v$ in 
Fourier space, with the vanishing sound speed $c_{\rm m}$.
For the perturbed line element (\ref{permet}), 
their linear perturbation equations of motion 
are given by 
\ba
& &
\dot{\delta \rho}_{\rm m}+3H \delta \rho_{\rm m}-3\rho_{\rm m}\dot{\Phi}
+\frac{k^2}{a^2} \rho_{\rm m} v=0\,,
\label{perma1}\\
& &
\dot{v}-\Psi=0\,.
\label{perma2}
\ea
Taking the time derivative of Eqn.~(\ref{perma1}) 
and using Eqn.~(\ref{perma2}), the matter 
density contrast 
$\delta_{\rm m}=\delta \rho_{\rm m}/
\rho_{\rm m}$ obeys
\be
\ddot{\delta}_{\rm m}+2H \dot{\delta}_{\rm m}+\frac{k^2}{a^2} \Psi=
3\ddot{\Phi}+6H \dot{\Phi}\,.
\label{delmse}
\ee
In order to derive an explicit relation between 
$\Psi$ and $\delta_{\rm m}$, we need to resort to a so called quasi-static 
approximation (QSA) \cite{Boisseau:2000pr,Tsujikawa:2007gd,DeFelice:2011hq,Sawicki:2015zya}. 
For the modes inside the sound horizon, this amounts to picking up 
the terms containing $\delta_{\rm m}$ 
and $k^2/a^2$ in the perturbation 
equations of motion arising from 
the gravity sector\footnote{Within the Horndeski class of models, 
it has been proven to be a valid assumption for the wavenumber 
in the range $k > 10^{-3}~h$/Mpc \cite{Peirone:2017ywi, Frusciante:2018jzw}.}. 
The QSA was first exploited in full 
Horndeski theories in Ref.~\cite{DeFelice:2011hq}.
Under this approximation, $\Psi$ and $\Phi_{\rm eff}$ 
are related to $\delta_{\rm m}$ through Eqs.~(\ref{mudef}) and 
(\ref{sigmadef}), respectively, with \cite{Kase:2018aps} 
\ba
\mu=\Sigma=1+\frac{4\dot{\phi}^4 G_{3,X}^2}{q_s c_s^2}\,,
\label{muSigma}
\ea
where 
\ba
q_s &=& 4M_{\rm pl}^2 \left(-G_{2,X}+2\dot{\phi}^2
G_{2,XX}+6 H \dot{\phi} G_{3,X}
-6 H \dot{\phi}^3 G_{3,XX} 
\right)+12 \dot{\phi}^4 G_{3,X}^2\,,
\label{qsg}\\
c_s^2 &=& \frac{-4M_{\rm pl}^2 (G_{2,X}-2\ddot{\phi} 
G_{3,X}-4H \dot{\phi}G_{3,X}+2\ddot{\phi}
\dot{\phi}^2 G_{3,XX})-4\dot{\phi}^4 
G_{3,X}^2}{q_s}\,.
\label{csg}
\ea
Under the QSA, the right hand-side of 
Eqn.~(\ref{delmse}) is negligible relative to 
its left hand-side, so using Eqn.~(\ref{mudef}) leads to 
\be
\ddot{\delta}_{\rm m}+
2H \dot{\delta}_{\rm m}
-4\pi G_{\rm N} \mu \rho_{\rm m} 
\delta _{\rm m}
\simeq 0\,.
\label{delmse2}
\ee

The absence of scalar ghosts requires that 
\be
q_s>0\,.
\label{qs}
\ee
The quantity $c_s^2$ corresponds to the scalar sound speed squared in the sub-horizon limit. The Laplacian instability is absent 
for
\be
c_s^2 > 0\,.
\label{cs}
\ee
Under the conditions (\ref{qs}) and 
(\ref{cs}), both $\mu$ and $\Sigma$ are larger than 1. 
Hence the gravitational interaction for 
linear perturbations inside the sound 
horizon ($c_s k >aH$) is stronger than that in the 
$\Lambda$CDM model. 
For the models of cosmic acceleration in which 
the cubic derivative interaction $G_3(X)$ contributes to the 
dark energy density, $\Sigma$ grows 
at low redshifts. In such cases the 
condition (\ref{Sigmad}) is satisfied, 
so there is a possibility for realizing 
the negative ISW-galaxy cross-correlation.

The property $\mu=\Sigma$ holds for the theories 
given by the Lagrangian $L=G_2(\phi,X)+G_3(\phi,X)
\square \phi+(M_{\rm pl}^2/2)R$, but 
$\Sigma$ is different from $\mu$ 
in the presence of a nonminimal coupling $G_4(\phi)R$ \cite{Kase:2018aps}.
For example, the theories with the Lagrangian 
$L=G_2(\phi, X)+G_4(\phi)R$, which include 
Brans-Dicke theories and $f(R)$ gravity 
as specific cases, lead to 
$\Sigma=M_{\rm pl}^2/[2G_4(\phi)]$. 
If the scalar field does not vary much at late times, which is typically the case for dark energy models in the framework 
of Brans-Dicke theories and $f(R)$ gravity, 
the variation of $\Sigma$ is not large enough to 
lead to the negative ISW-galaxy cross-correlation.
As we will see in Sec.~\ref{K-mouflage_model}, 
K-mouflage belongs to this latter subclass of Horndeski theories.

\subsubsection{Galileon Ghost Condensate (GGC)} \label{GGC_model}

As a first example within the framework of the action (\ref{actionHo}), we consider the GGC model characterized by the functions \cite{Deffayet:2010qz,Kase:2018iwp}
\be
G_2(X)=a_1 X+a_2 X^2\,,\qquad 
G_3(X)=3a_3X\,,
\label{actionGGC}
\ee
where $a_{1,2,3}$ are constants.
This model generalizes the Cubic Galileon model (hereafter G3)~\cite{Deffayet:2009wt} by taking the term $a_2X^2$ into account, 
which modifies the cosmic expansion and growth histories of both
linear \cite{Kase:2018iwp,Peirone:2019aua} and non-linear \cite{Frusciante:2020zfs} perturbations compared to G3. 
In particular, it works to suppress the large-scale CMB temperature 
anisotropy in comparison to the $\Lambda$CDM model.
These features lead to the statistical preference of GGC over $\Lambda$CDM 
when data from CMB, BAO, SNIa, and RSDs are used \cite{Peirone:2019aua}. Furthermore, the estimation of today's Hubble parameter $H_0$ constrained from CMB temperature and polarization data is consistent with its direct measurement at $2\sigma$, alleviating 
the Hubble tension \cite{Peirone:2019aua}.

We introduce dimensionless variables to compute some relevant quantities for our analysis, which are defined by
\be
x_1=-\f{a_1\dot{\phi}^2}{3M_{\rm pl}^2 H^2}\,,\qquad x_2=\f{a_2\dot{\phi}^4}{M_{\rm pl}^2 H^2}\,, \qquad x_3=\f{6a_3\dot{\phi}^3}{M_{\rm pl}^2H}
\,.
\label{dimensionless_functions}
\ee
The Friedmann equation is given by Eqn.~(\ref{back1}) 
with the dark energy 
density parameter 
\be
\Omega_{\rm DE}=x_1+x_2+x_3\,.
\ee
Solving Eqs.~(\ref{back2}) and 
(\ref{back3}) for 
$h \equiv \dot{H}/H^2$ and $\epsilon_{\phi} \equiv \dot{\phi}/(H \phi)$, 
it follows that 
\ba
h &=& -\frac{12x_1^2+2(16x_2+12x_3
+3\Omega_{\rm m}+4 \Omega_{\rm r})x_1+16 x_2^2+4(6x_3+3\Omega_{\rm m}+4\Omega_{\rm r})x_2+(9x_3+6\Omega_{\rm m}+8 \Omega_{\rm r})x_3}{4x_1+8x_2+4x_3+x_3^2}\,,
\label{hex}\\
\epsilon_{\phi} &=& 
\frac{-12 x_1-8x_2+(6x_1+4x_2+3\Omega_{\rm m}+4\Omega_{\rm r}-6)x_3+3x_3^2}{4x_1+8x_2+4x_3+x_3^2}\,.
\label{epex}
\ea
The dark energy equation of state 
(\ref{wde}) yields
\be
w_{\rm DE}=\frac{3x_1+x_2-\epsilon_{\phi}
x_3}{3(x_1+x_2+x_3)}\,.
\label{wdef}
\ee
Let us consider the case in which the cubic derivative 
interaction $G_3(X)$ dominates over $G_2(X)$ in the radiation- or 
matter-dominated epochs. 
Since the condition 
$\{ |x_1|, |x_2|\} \ll |x_3| \ll 1$ is satisfied in this regime, Eqs.~(\ref{hex}) and (\ref{epex}) give the approximate relations
\be
h \simeq -\frac{1}{2} (3\Omega_{\rm m}
+4\Omega_{\rm r})\,,\qquad 
\epsilon_{\phi} \simeq 
\frac{1}{4} (3\Omega_{\rm m}+4\Omega_{\rm r}
-6)\,.
\ee
Then, from Eqn.~(\ref{wdef}), we obtain
\be
w_{\rm DE} \simeq -\frac{1}{3}\epsilon_{\phi} \simeq \frac{1}{2}-\frac{1}{4}\Omega_{\rm m}
-\frac{1}{3} \Omega_{\rm r}\,,
\label{wdeEar}
\ee
so that $w_{\rm DE} \simeq 1/6$ in the radiation era 
and $w_{\rm DE} \simeq 1/4$ 
in the matter era.

Since $|x_1|$ increases faster than $|x_3|$, 
it can happen that the solutions approach 
a tracker solution satisfying 
$H\dot{\phi}={\rm constant}$. 
This condition translates to $h=-\epsilon_{\phi}$, 
so that $x_3=-2x_1$ for $x_2 \to 0$.
Ignoring $x_2$ in comparison to $x_1$ and 
$x_3$, the dark energy equation of state 
for the tracker yields 
\be
w_{\rm DE}=-1+\frac{2}{3}h\,,
\ee
so that $w_{\rm DE}=-7/3$ during the 
radiation era and $w_{\rm DE}=-2$ 
during the matter era.  
The moment at which the solutions 
approach the tracker depends on the initial 
conditions of $x_1$, $x_2$, and $x_3$. 
The existence of nonvanishing $x_2$ works to 
prevent the solutions from approaching 
the tracker.
Even when the tracker solution is not reached, 
it is possible to realize the phantom dark energy 
equation of state ($w_{\rm DE}<-1$) 
without having ghosts \cite{Kase:2018iwp,Peirone:2019aua}. 
Finally, the cosmological trajectories 
converge to a de Sitter attractor 
($w_{\rm DE}=-1$, $\Omega_{\rm DE}=1$, 
$\Omega_{\rm m}=0$, $\Omega_{\rm r}=0$), 
at which the two relations 
$x_1=-2+x_3/2$ and $x_2=3-3x_3/2$ hold.

{}From Eqs.~(\ref{qsg}) and (\ref{csg}), 
the conditions for the absence of ghosts 
and Laplacian instabilities are given, respectively, by
\ba
q_s &=& \frac{3H^2 M_{\rm pl}^4}{\dot{\phi}^2}
(4x_1+8x_2+4x_3+x_3^2)>0\,,
\label{qscon}\\
c_s^2 &=& \frac{12x_1+8x_2
+4(\epsilon_{\phi}+2)x_3-x_3^2}
{3(4x_1+8x_2+4x_3+x_3^2)}>0\,.
\label{cscon}
\ea
In the regime characterized by  
$\{ |x_1|, |x_2|\} \ll |x_3| \ll 
1$, these two conditions are satisfied for 
$x_3>0$. On the tracker ($x_3=-2x_1$ 
with $x_2 \to 0$), we require that $x_1<0$ for the consistency 
with (\ref{qscon}) and (\ref{cscon}). 
On the late-time de Sitter solution, the 
quantities (\ref{qscon}) and (\ref{cscon}) 
reduce, respectively, to 
\be
q_s=\frac{4H^2 M_{\rm pl}^4}{3\dot{\phi}^2} 
\left( x_2^2+3x_2+18 \right)\,,\qquad
c_s^2=\frac{x_2 (3-x_2)}{3(x_2^2+3x_2+18)}\,.
\ee
Hence we require the condition $0<x_2<3$ 
for the stability of the de Sitter solution.
We will consider the cosmological evolution 
in which $\dot{\phi}$ does not change 
the sign. Then, the signs of $x_1$, $x_2$, 
and $x_3$ consistent with the stability 
conditions are 
\be
x_1<0\,,\qquad x_2>0\,,
\qquad x_3>0\,.
\ee
Today's values of $x_1$, $x_2$, and $x_3$ 
constrained by the observational data of 
CMB, BAO, SNIa, and RSD data do exist 
in this range \cite{Peirone:2019aua}.

{}From Eqn.~(\ref{muSigma}), we have
\be
\mu=\Sigma=1+\frac{x_3^2}
{12 x_1+8x_2+4(\epsilon_{\phi}+2)x_3
-x_3^2}\,.
\label{muSigGa}
\ee
In the two early-time cosmological epochs 
discussed above, the right hand-side of 
Eqn.~(\ref{muSigGa}) reduces to 
\begin{equation}
 \mu=\Sigma \simeq
  \begin{cases}
    1+\dfrac{x_3}{2+3\Omega_{\rm m}+4\Omega_{\rm r}} &~~~{\rm for}~~
    \{ |x_1|, |x_2|\} \ll |x_3| \ll 1, \\
    \\
    1+\dfrac{x_3}{2(1+3\Omega_{\rm m}+4\Omega_{\rm r})} &~~~ {\rm for}~~x_3=-2x_1\,,\quad x_2 \to 0\,.
  \end{cases}
\end{equation}
With the growth of $x_3$, $\Sigma$ increases in time. 
The growth of $\Sigma$ is particularly significant in the late 
universe at which $x_3$ contributes to 
the dark energy density.
Hence it is possible to realize the negative 
ISW-galaxy cross-correlations in the GGC model.

\subsubsection{Generalized Cubic Covariant Galileon (GCCG)} 
\label{GCCG_model}

Let us consider the GCCG model~\cite{DeFelice:2011bh} 
specified by the functions
\be
G_2(X)=-c_2 M_2^{4(1-p)} (-X/2)^{p}\,,
\qquad
G_3(X)=-c_3 M_3^{1-4p_3} (-X/2)^{p_3}\,,
\ee
where $c_2$, $c_3$, $p$, $p_3$ are dimensionless constants, 
and $M_2$, $M_3$ 
are constants having a dimensionless of mass.
Instead of $p_3$, we will use the parameter $q$ defined by 
\be
q=p_3-p+\frac{1}{2}\,.
\ee
The GCCG model allows the existence of a tracker solution obeying the relation 
$H \dot{\phi}^{2q}={\rm constant}$ with 
$q>0$. We note that the G3 model corresponds 
to the powers $p=1$, $p_3=1$, and 
$q=1/2$. In the following, we will consider 
the case of positive values of $p$ and $q$, which are actually 
consistent with the stability 
conditions discussed below.

To study the cosmological dynamics, it is 
convenient to introduce the following dimensionless variables \cite{DeFelice:2011bh,DeFelice:2011aa}
\be
x \equiv \frac{\dot{\phi}}{HM_{\rm pl}}\,,
\qquad
r_1 \equiv \left( \frac{x_{\rm dS}}{x} 
\right)^{2q} \left( \frac{H_{\rm dS}}{H} 
\right)^{1+2q}\,,
\qquad
r_2 \equiv \left[ \left( \frac{x}{x_{\rm dS}}
\right)^2 \frac{1}{r_1^3} \right]^{\frac{p+2q}{1+2q}}\,,
\ee
where $x_{\rm dS}$ and $H_{\rm dS}$ are 
the values of $x$ and $H$ on the late-time 
self-accelerating de Sitter solution, 
respectively (at which $\dot{\phi}$ and 
$H$ are constants with $r_1=1$ and $r_2=1$). 
We will consider the solution in the range 
$x>0$ without loss of generality.

We relate the masses $M_2$ and $M_3$ 
with $H_{\rm dS}$, as
\be
M_2 = \left( H_{\rm dS} M_{\rm pl} 
\right)^{1/2}\,,\qquad 
M_3 = \left( H_{\rm dS}^{-2p_3} 
M_{\rm pl}^{1-2p_3} \right)^{1/(1-4p_3)}\,.
\ee
Using Eqs.~(\ref{back1}) and (\ref{back2}) on the 
de Sitter solution, the coefficients $c_2$ and $c_3$ 
are expressed as
\be
c_2=3 \cdot 2^p x_{\rm dS}^{-2p}\,,\qquad
c_3=\frac{2^{p+q+1/2}}{2p+2q-1}p
x_{\rm dS}^{-2(p+q)}\,.
\ee

The density parameter and equation of 
state of dark energy are given, 
respectively, by 
\ba
\Omega_{\rm DE} &=& 
\left[2p-(2p-1)r_1 
\right] r_1^{1+\alpha}r_2\,,\\
w_{\rm DE} &=& -\frac{(\Omega_{\rm r}-3+
12r_1-6r_1^2)p+3(2q-1)r_1+3r_1^2}
{3[(2-2r_1+r_1^{1+\alpha}r_2)p+2q-1+r_1]
[2p+(1-2p)r_1]}\,,
\label{wdegc}
\ea
where 
\be
\alpha \equiv \frac{p+2q}{1+2q}\,.
\ee
For the derivation of Eqn.~(\ref{wdegc}), 
we eliminated $\Omega_{\rm m}$ by using 
Eqn.~(\ref{Omecon}).

The variable $r_1$ obeys the 
differential equation 
\be
r_1'({\cal N}) = \frac{r_1(r_1-1)[3-6p-12q+(1-2p)
\Omega_{\rm r}-3r_1^{2+\alpha}r_2
-6p(1-r_1)r_1^{1+\alpha}r_2]}
{2p(2-2r_1+r_1^{1+\alpha}r_2)-2+4q+2r_1}\,.
\label{r1eq}
\ee
It is also straightforward to derive the 
differential equations for 
$r_2$ and $\Omega_{\rm r}$ \cite{DeFelice:2011bh,DeFelice:2011aa}.
In Eqn.~(\ref{r1eq}), there are two kinds of  fixed points characterized by $r_1=0$ and $r_1=1$. 
The tracker corresponds to 
the fixed point satisfying $r_1=1$ and 
$r_2 \ll 1$, whereas the de Sitter solution 
is characterized by $r_1=1$ and $r_2=1$.
For the initial condition
$r_1 \ll 1$, the quantity $r_1$ can grow to reach 
the tracker during the radiation or matter era. After the growth of $r_2$ 
toward 1, the solutions finally converge 
to the de Sitter fixed point.

{}From Eqn.~(\ref{wdegc}), the dark energy 
equation of state in the regime 
$r_1 \ll 1$ and $r_2 \ll 1$ is approximately
given by\footnote{In the presence of quatric 
and quintic Horndeski interactions, this asymptotic
value of $w_{\rm DE}$ 
is subject 
to modifications \cite{DeFelice:2011bh}.} 
\be
w_{\rm DE} \simeq \frac{3-\Omega_{\rm r}}{12(p+q)-6}\,.
\ee
For the G3 model ($p=1$ and $q=1/2$), this reduces 
to $w_{\rm DE} \simeq 
1/4-\Omega_{\rm r}/12$, 
whose value coincides with Eqn.~(\ref{wdeEar}) derived 
in the limit $\{ |x_1|, |x_2|\} \ll |x_3| \ll 1$ 
in the GGC model.

On the tracker characterized by $r_1=1$ 
and $r_2 \ll 1$, the dark energy equation 
of state is given by 
\be
w_{\rm DE}=-1-\frac{1}{6} (3+\Omega_{\rm r})s\,,
\ee
where 
\be
s \equiv \frac{p}{q}\,.
\label{sdef}
\ee
The quantity $s$ characterizes the deviation 
from the $\Lambda$CDM model\footnote{In Refs.~\cite{DeFelice:2011bh,DeFelice:2011aa}, 
the definition of $s$ is half of the 
right hand side of Eqn.~(\ref{sdef}).}. 
For positive values of $p$ and $q$, 
the phantom tracker equation 
of state can be realized. 
Finally, the solutions approach 
the de Sitter attractor ($r_1=r_2=1$) 
with $w_{\rm DE}=-1$. 

For the quantities associated with 
perturbations, we have 
\ba
\tilde{q}_s &\equiv& \frac{q_s}
{4\dot{\phi}^4 G_{3,X}^2}
=3+\frac{3}{pr_1^{1+\alpha}r_2} 
\left[ 2p+2q-1+(1-2p)r_1 \right]\,,\\
c_s^2 &=& [(2p+2q-1)\Omega_{\rm r}-5
+r_1 (8-6r_1+3r_1^{1+\alpha}r_2)
+2p\{5-8r_1+6r_1^2+(5-2q-7r_1)
r_1^{1+\alpha}r_2 \}+2q (5-3r_1^{2+\alpha}r_2) \nonumber\\
& &-2p^2 r_1^{1+\alpha}r_2 (2-2r_1+ r_1^{1+\alpha}
r_2)]/[6\{(2-2r_1+r_1^{1+\alpha}r_2)p+2q-1+r_1\}^2]\,,\\
\mu &=& \Sigma=
1+\frac{1}{\tilde{q}_s c_s^2}\,.
\ea
When $r_1 \ll 1$ and 
$r_2 \ll 1$, the conditions 
$\tilde{q}_s>0$ and $c_s^2>0$ 
are satisfied for 
\be
2p+2q-1>0\,.
\ee
In this regime, expanding $\mu$ around $r_1=0$ and $r_2=0$ 
leads to 
\be
\mu=\Sigma \simeq 1+\frac{2p}{5+\Omega_{\rm r}}
r_1^{1+\alpha}r_2\,,
\ee
whose deviation from 1 is suppressed to be small. 

On the tracker with $r_1=1$ and $r_2 \ll 1$,  the stability conditions reduce to 
\ba
\tilde{q}_s &=&
3+\frac{6q}{pr_2}>0\,,\\
c_s^2 &=& \frac{6p+10q-3+(2p+2q-1)\Omega_{\rm r}}
{24q^2}>0\,,
\ea
which constrain the allowed parameter regions 
of $p$ and $q$.
Expansion of $\mu$ around $r_2=0$ gives 
\be
\mu=\Sigma \simeq 1+\frac{4pq}
{6p+10q-3+(2p+2q-1)\Omega_{\rm r}}r_2\,.
\label{musitra}
\ee
As the variable $r_2$ grows toward the de Sitter fixed point, 
the deviation of $\Sigma$ from 1 tends to be larger.

At $r_1=r_2=1$, the stability conditions translate to 
\ba
\tilde{q}_s &=& \frac{3(p+2q)}{p}>0\,,\\
c_s^2 &=& \frac{1-p}{3(p+2q)}>0\,.
\ea
For $p>0$, we require that 
$p+2q>0$ and $p<1$.
On the de Sitter fixed point, we also have 
\be
\mu=\Sigma=\frac{1}{1-p}\,,
\ee
which is larger than 1 for 
$0<p<1$.

For the data analysis, we will exploit the two parameters 
$s$ and $q$ instead of $p$ and $q$. 
The observational constraints on the GCCG model were carried out 
with the data of CMB, BAO, and SNIa \cite{DeFelice:2011aa}, 
showing that nonvanishing positive values of $s$ are favored. 
The analysis with the Planck CMB data alone implies that the constrained value of $H_0$ is consistent with its determination from Cepheids 
at 1$\sigma$ \cite{Frusciante:2019puu}.
Moreover, it is possible to realize either positive or negative 
ISW-galaxy cross-correlations \cite{DeFelice:2011aa,Giacomello:2018jfi}, depending on the model parameters.

\subsection{K-mouflage} 
\label{K-mouflage_model}

We also consider K-mouflage theories in which 
the action in the Einstein frame is described by  
\cite{Brax:2014wla,Brax:2015dma,Brax:2016Km}
\be
{\cal S} = \int {\rm d}^4 x \sqrt{-\tilde{g}} \left[ \frac{M_{\rm pl}^2}{2} \tilde{R} + {\cal M}^4 K(\tilde{\chi}) \right]  
+{\cal S}_{\rm m}(\psi_{\rm m}, g_{\mu\nu})
\label{eq:actkm}
\ee
where ${\cal M}^4$ is some energy scale, and $K$ is 
a function of the dimensionless field kinetic 
energy given by 
\be
\tilde{\chi} = 
- \frac{\tilde{g}^{\mu\nu} \nabla_{\mu}\phi \nabla_{\nu}\phi}
{2 {\cal M}^4} \,.
\ee
We use a tilde to represent quantities in the Einstein 
frame. The metric $g_{\mu \nu}$ in the Jordan frame is 
related to $\tilde{g}_{\mu \nu}$ through 
a $\phi$-dependent conformal factor $A(\phi)$, as
\be
g_{\mu \nu}=A^2(\phi)\tilde{g}_{\mu \nu}\,.
\label{gmutra}
\ee
The action ${\cal S}_{{\rm m}}$ depends on the matter fields 
$\psi_{m}$ and the Jordan-frame metric $g_{\mu \nu}$. 
The matter fields in the Einstein frame 
(metric $\tilde{g}_{\mu \nu}$) are directly 
coupled to the scalar field $\phi$ 
through the conformal factor $A(\phi)$.
K-mouflage theories are endowed with a screening mechanism 
acting in regions where the kinetic term becomes non-linear, 
which can suppress fifth forces 
induced by the coupling function $A(\phi)$. 
This typically happens on small cosmological scales, 
when the first derivative of the scalar field is large enough.

Let us consider the Jordan-frame action in the form 
\be
{\cal S}=\int {\rm d}^4 x \sqrt{-g} \left[ 
G_4(\phi)R+G_2(\phi,X) \right]
+{\cal S}_{\rm m} (\psi_{\rm m}, g_{\mu \nu})\,,
\label{actionHo2}
\ee
which corresponds to a nonminimally coupled k-essence. 
The Ricci scalar $\tilde{R}$ in the Einstein frame 
is related to $R$ according to 
$R=\tilde{R}+6 \tilde{\square}\omega
-6 \tilde{g}^{\mu \nu} \partial_{\mu}\omega 
\partial_{\nu} \omega$, where $\omega=-\ln A$. 
In order to recast the action (\ref{actionHo2}) 
to that in the Einstein frame 
under the transformation (\ref{gmutra}), 
the conformal factor $A(\phi)$ is related to  $G_4(\phi)$, as
\be
G_4(\phi)=\frac{M_{\rm pl}^2}{2A^2(\phi)}\,.
\ee
Today's value of $A(\phi_0)$ is normalized to be 1 
to recover the gravitational coupling 
$G_4(\phi_0)=M_{\rm pl}^2/2$.
Then, the Einstein-frame action following from 
Eqn.~(\ref{actionHo2}) is of the form 
\be
{\cal S} = \int {\rm d}^4 x \sqrt{-\tilde{g}} 
\left( \frac{M_{\rm pl}^2}{2} \tilde{R} 
+\frac{6M_{\rm pl}^2 {\cal M}^4 A_{,\phi}^2}{A^2} \tilde{\chi}+A^4 G_2
\right)+{\cal S}_{\rm m}(\psi_{\rm m}, g_{\mu\nu})\,,
\label{actionEi}
\ee
up to a boundary term. 
Here, we used the fact that the field kinetic energy 
$X=g^{\mu \nu} \nabla_{\mu} \phi \nabla_{\nu} \phi$ 
is related to $\tilde{\chi}$ as 
\be
\tilde{\chi}=-\frac{A^2(\phi)X}{2{\cal M}^4}\,.
\ee
Comparing (\ref{eq:actkm}) with (\ref{actionEi}), 
there is the following correspondence
\be
G_2(\phi,X)=\frac{{\cal M}^4}{A^4(\phi)}
K (\tilde{\chi})-\frac{6{\cal M}^4 M_{\rm pl}^2
A_{,\phi}^2(\phi)} {A^6 (\phi)} 
\tilde{\chi}\,.
\label{G2pX}
\ee

The kinetic function $K(\tilde{\chi})$ and  
nonminimal coupling $A(\phi)$ are the two 
key quantities to determine 
cosmic expansion and growth histories. 
We will study the cosmological dynamics of K-mouflage
in the Jordan frame. 
In doing so, it is convenient to 
introduce the following dimensionless quantities
\be
\epsilon_1 \equiv \frac{2M_{\rm pl}^2A_{,\phi}^2}
{K_{,\tilde{\chi}}A^2}\,,
\qquad
\epsilon_2 \equiv \frac{A_{,\phi}\dot{\phi}}{HA}\,.
\label{epsilon2-def}
\ee
In the Jordan frame, the Friedmann equation 
and scalar-field equation are given, 
respectively, by 
\ba 
& &
3 M_{\rm pl}^2 H^2 (1-\epsilon_2)^2=
A^2(\rho_{\rm m}+\rho_{\rm r})+
\frac{{\cal M}^4}{A^2} \left( 
2\tilde{\chi} K_{,\tilde{\chi}}-K \right) \,,
\label{E00}\\
& & 
\frac{{\rm d} }{{\rm d} t} \left( A^{-2}a^3 \dot{\phi} 
K_{,\tilde{\chi}} \right)=-a^3 \rho_{\rm m}
\frac{A_{,\phi}}{A}\,.
\label{E00d}
\ea
The term appearing on the right hand side 
of Eqn.~(\ref{E00d}) is proportional to a trace of the
matter energy-momentum tensor, so the contribution 
of radiation to it vanishes.

On using the general stability criteria derived for 
full Horndeski theories \cite{Kase:2018aps}, 
the conditions for the absence of scalar 
ghosts and Laplacian instabilities 
translate, respectively, to 
\ba
q_s &=& \frac{2M_{\rm pl}^2}{A^4} 
(K_{,\tilde{\chi}}+2\tilde{\chi}
K_{,\tilde{\chi \chi}})>0\,,\label{qsK}\\
c_s^2 &=& \frac{K_{,\tilde{\chi}}}{K_{,\tilde{\chi}}+2\tilde{\chi}
K_{,\tilde{\chi \chi}}}>0\,,\label{csK}
\ea
which will be imposed as theoretical 
priors. The evolution of linear perturbations 
in the Jordan frame is modified by the presence 
of nonminimal coupling $A(\phi)$. 
Employing the QSA and neglecting the mass term arising 
from $G_{2,\phi \phi}$, the dynamics of gravitational 
potentials is governed by 
Eqs.~\eqref{mudef}-\eqref{sigmadef}, 
with \cite{Benevento:2018xcu}
\be
\mu=(1+ \epsilon_1)A^2 \, , 
\qquad \Sigma=A^2 \,, 
\label{mu_Sigma_KM}
\ee
so that $\mu \neq \Sigma$. 
The time variation of nonminimal coupling $A(\phi)$ 
induced by the background evolution of $\phi$ affects 
the observable associated with the ISW-galaxy cross-correlation.

To study the dark energy dynamics in K-mouflage 
theories, we need to specify the functional 
forms of $K(\tilde\chi)$ and $A(\phi)$. 
In our analysis, we exploit a parameterization introduced in Ref.~\cite{Brax:2016Km}, where the time dependence of $A$ and $K$ is determined by five constant parameters: $\{\epsilon_{2,0},\gamma_A, m, \alpha_U,\gamma_U\}$. 
The function $A(a)$ is then expressed as 
\be
A(a)=1+\alpha_A- \alpha_A \left[ 
\frac{a (\gamma_A+1)}{a+\gamma_A} \right]^{\nu_A} 
\label{A_def}\,,
\ee
where
\be
\nu_A=\frac{3(m-1)}{2m-1} \,,
\qquad
\alpha_A=
-\frac{\epsilon_{2,0}(\gamma_A+1)}{\gamma_A \nu_A} \,,
\ee
where $\epsilon_{2,0}$ is today's value of $\epsilon_{2}$.
The $\alpha_U$ and $\gamma_U$ parameters
contribute to the determination of $K(a)$, 
which can be computed by the solution of the following differential equation
\begin{align}
\frac{\mathrm{d} K}{\mathrm{d} \tilde{\chi}} &=
\frac{U(a)}{a^3 \sqrt{\tilde{\chi}}} 
\,
\label{eq:dK/dchi}, 
\end{align}
with
\begin{align}
U(a) &= U_0 \left[ \sqrt{a_{\mathrm{eq}}}+1+ \frac{\alpha_U}{\ln(\gamma_U+1)} \right] \frac{a^2 \ln(\gamma_U+a)}{(\sqrt{a_{\mathrm{eq}}}+\sqrt{a}) \ln(\gamma_U+a)+ \alpha_U a^2} \label{U_def}  \,, \\
\sqrt{\tilde{\chi}}& =-\frac{\rho_{{\rm m},0}}{{\cal M}^4} 
\frac{\epsilon_2 A^4}{2U[({\rm d} \ln U/{\rm d} \ln a)
-3 \epsilon_2]}\,,\label{chi_tilde_U} 
\end{align}
where $a_{\rm eq}$ is the scale factor at the epoch of 
matter-radiation equality, and $\rho_{{\rm m},0}$ is 
today's value of $\rho_{\rm m}$.

The theoretically allowed parameter space in K-mouflage theories 
were investigated in Refs.~\cite{Brax:2014wla,Brax:2016Km}. 
For $A>0$ and $A_{,\phi}>0$ together with the stability conditions 
(\ref{qsK}) and (\ref{csK}), the time derivative of $\phi$ being 
compatible with Eq.~(\ref{E00d}) should be in the range 
$\dot{\phi}<0$. This translates to the inequality $\epsilon_2<0$.
Taking the ${\cal N}=\ln a$ derivative of $\Sigma=A^2$, 
it follows that 
\be
\Sigma'({\cal N})=\epsilon_2 A^2\,.
\ee
Since $\Sigma'({\cal N})<0$ under the condition $\epsilon_2<0$, 
the necessary condition (\ref{Sigmad}) for realizing the negative 
ISW-galaxy cross-correlation is not satisfied in K-mouflage theories. 
This means that K-mouflage theories always give rise to a 
positive ISW-galaxy cross-correlation, whose property 
is different from those in the GGC and GCCG models.

Cosmological constraints up to linear scales using CMB, SNIa, BAO, 
and local measurements of $H_0$ set a lower $2\sigma$ limit of $- 0.04$ 
for $\epsilon_{2,0}$, while the other parameters 
are unconstrained \cite{Benevento:2018xcu}.
This is because at leading order $\epsilon_1 \simeq - \epsilon_2$ 
during the matter era, so most of the deviations from 
$\Lambda$CDM are determined by the value $\epsilon_{2,0}$. 
The $\epsilon_{2,0}$ parameter also impacts the ISW-galaxy cross-correlation in a two-fold way. On one hand, a higher value of $\epsilon_{2,0}$ leads to 
a faster cosmic expansion at late times, 
whose effect works to suppress
the growth of matter perturbations.
On the other hand, the effective dimensionless gravitational coupling 
$\mu=(1+\epsilon_1)A^2$ is enhanced by the positive term $\epsilon_1$ 
induced by the $\phi$-dependent coupling $A(\phi)$. 
The trade-off between these two effects, together with the decrease of 
$\Sigma$ in time, leads to a higher amplitude of 
the ISW effect in K-mouflage theories compared to its 
$\Lambda$CDM-limit ($\epsilon_{2,0} \rightarrow 0$).

\section{Methodology and Data}\label{Sec:method}

\subsection{Einstein-Boltzmann code}

To perform the likelihood analysis of finding the best-fit parameters 
of each model discused in Sec.~\ref{Sec:models}, 
we use the publicly available  Einstein-Boltzmann code \eftcamb\ \cite{Hu:2013twa,Raveri:2014cka,Hu:2014oga}~\footnote{Web page: \url{http://www.eftcamb.org}} developed to study dark energy and MG models based on the Effective Field theory (EFT) of 
dark energy~\cite{Gubitosi:2012hu,Bloomfield:2012ff} 
(see \cite{Kase:2014cwa,Frusciante:2019xia} for reviews). 
\eftcamb \, includes \eftcamb\_\texttt{sources}, based on \camb\_\texttt{sources} \cite{Challinor:2011bk}, 
which is useful to extract theoretical predictions for galaxy number 
counts and cross-correlations with CMB. 
The specific models we investigate in this paper can be encoded 
in the EFT formalism, given the possibility to map Lagrangian-based 
theories with an additional scalar degree of freedom into the EFT approach~\cite{Gubitosi:2012hu,Bloomfield:2012ff,Bloomfield:2013efa,Gleyzes:2013ooa,Gleyzes:2014rba,Frusciante:2015maa,Frusciante:2016xoj}. 
In particular, we refer the reader to Ref.~\cite{Peirone:2019aua,Peirone:2019yjs} 
for the mapping and implementation of the GGC model, Ref.~\cite{Frusciante:2019puu} for GCCG and 
Ref.~\cite{Benevento:2018xcu} for K-mouflage. 
Additionally, we have implemented massive neutrinos in the EFT modules for GGC and K-mouflage. The GCCG module in \eftcamb\ already comes with massive neutrinos.

\subsection{The significance of the ISW}\label{Sec:ISWsignificance}

We illustrate the method for quantifying the significance 
of the ISW effect. 
To this end, we apply the likelihood code developed 
in Ref.~\cite{Stolzner:2017ged} to MG models. 
After presenting the data sets to be exploited 
in our analysis, we illustrate how we constrain the galaxy bias, which enters in Eqn.~\eqref{CLgg} via Eqn.~\eqref{Ig_def}. 
Finally, we introduce the likelihood used to provide measurements of 
the ISW effect through ISW-galaxy cross-correlations.

We implement cross-correlations of the CMB at large angular scales 
(multipoles $l \lsim 100$) from Planck 2015 temperature data \citep{planck_collaboration_planck_2016-1} \footnote{Planck 2018 temperature data are more up to date, but the differences between 2015 and 2018 releases were primarily found in improvements to polarization. 
Hence it is unlikely that the differences between two data sets for 
$l \lsim 100$ lead to a significant change \citep{Aghanim:2018eyx}.}, 
with  galaxy number counts from: the 2Mass Photometric Catalog (2MPZ) \cite{Bilicki/etal:2014,Alonso/etal:2015}, 
WISE x SuperCOSMOS photo-$z$ catalog (WIxSC) \cite{Bilicki/etal:2016}, 
the Sloan Digital Sky Survey Data Release 12 (SDSS-DR12) photo-$z$ sample \cite{Beck/etal:2016}, photometric quasars (QSO) from SDSS DR6 \cite{Richards/etal:2009}, and the National Radio Astronomy Observatory Very Large Array Sky Survey (NVSS) of extragalactic radio 
sources \cite{Condon/etal:1998}. 

These collections of data sets provide a tracer of 
the large-scale structure over a redshift range 
$0 < z < 5$. Cross-correlations between catalogs have been removed using appropriate masks, see Ref. \cite{Stolzner:2017ged} for a detailed description. For each galaxy catalog, we use a selection function defined as a convolution of the redshift distribution of galaxies, $\rm{d}N/\rm{d}z$, and a photometric redshift error function $p(z)$. As the redshift distribution functions, we exploit the publicly available distributions obtained from 
Ref.~\cite{Stolzner:2017ged}. 
For the photometric error functions, we adopt Gaussian distributions with errors defined for each catalog. For 2MPZ and QSO, we use photometric redshift errors given by $\sigma_z = 0.015$ and $\sigma_z = 0.24$, respectively. We implement redshift dependent photometric redshift errors 
for SDSS and WIxSC given by 
$\sigma_z = 0.022(1 + z)$ and $\sigma_z = 0.033(1 + z)$, 
respectively. For NVSS, we do not convolve the redshift distribution function 
with a photometric redshift error function as there is only one redshift bin. 

In our analysis, we fix the model and cosmological parameters to the best-fit ones obtained with a combination of CMB, BAO, SNIa data 
as described in Sec.~\ref{sec:bestfit} in detail. 
These are then exploited to compute the expected galaxy 
auto-correlation $\hat{C}_{l}^{\rm gg}$ 
[see Eqn.~(\ref{CLgg})] for each catalog 
and redshift bin. To fit the linear bias $b$, 
we compute the $\chi^2$ defined as
\be
\chi^2_{b} = \chi^2(b^2) = \displaystyle
\sum_{l{\rm -bins}} 
\frac{[\hat{C}_{l}^{\rm gg}(b^2) 
- C_{l}^{\rm gg}]^2}{(\Delta C_{l}^{\rm gg})^2}\,,
\label{auto_chi}
\ee
where $\hat{C}_{l}^{\rm gg}(b^2)$ 
assumes some value of the galaxy bias, 
$C_{l}^{\rm gg}$ is the measured auto-correlation 
power spectrum, and $(\Delta C_{l}^{\rm gg})^2$ 
is the data covariance matrix. 
The sum is over all $l$ bins. 
The above relation assumes that there is a unique 
galaxy bias factor for each redshift bin of each catalog.

At this point, along with the best-fit values for the model 
and cosmological parameters, we also fix 
the galaxy biases to their best-fit values. 
We then compute the ISW-galaxy cross-correlation likelihood 
using a $\chi^2$ given by
\be\label{eq:chi2cc}
\chi^2_{A_{\rm ISW}} = \chi^2(A_{\rm ISW}) = \displaystyle\sum_{\rm catg.} \displaystyle 
\sum_{z-{\rm bins}} \displaystyle\sum_{l{\rm -bins}} 
\frac{( A_{\rm ISW}\hat{C}_{l}^{\rm Tg} - C_{l}^{\rm Tg})^2}
{(\Delta C_{l}^{\rm Tg})^2},
\ee
where $C_{l}^{\rm Tg}$ is the measured cross-correlation power spectrum,  
$\hat{C}_{l}^{\rm Tg}$ is the theory cross-correlation power spectrum obtained with the best-fit parameters, 
and $(\Delta C_{l}^{\rm Tg})^2$ is the data 
covariance matrix. 
The only free parameter in the above relation is the amplitude $A_{\rm ISW}$. 
This is then the parameter for which we fit to data. $A_{\rm ISW}$ is assumed to take on a fiducial 
value of 1. When $A_{\rm ISW} \neq 1$, it rescales the amplitude of the theoretical ISW-galaxy cross-correlation power spectrum. In detail, $A_{\rm ISW} > 1$ corresponds to the theory predicting a lower ISW power relative to the measured one, while the opposite holds 
for $A_{\rm ISW} < 1$.

We compute the mean and standard deviation for 
$A_{\rm ISW}$ and the galaxy bias parameters 
using the following method. 
For a general parameter $B$ that rescales a theory 
vector $\bm{t}$ relative to some data vector $\bm{d}$ 
and some data covariance matrix $\mathbb{C}$, 
the best-fit value can be analytically derived  
by minimizing the $\chi^2$ with 
respect to $B$ finding
\be
B_{\rm bf}=\frac{\bm{d}^T\mathbb{C}^{-1}\bm{t}}
{\bm{t}^T\mathbb{C}^{-1}\bm{t}}\,,
\ee
with variance given by
\be
\sigma_{B}^2 = 
\frac{1}{\bm{d}^T\mathbb{C}^{-1}\bm{t}}\,.
\ee
Assuming that the amplitude parameter distribution is approximately Gaussian, the mean value of $B$ is approximately equivalent to the best-fit value of $B$. Since both $A_{\rm ISW}$ and the galaxy bias parameters rescale the amplitude of the cross- and auto-correlation power spectra, respectively, it follows that this analytic formulation can be applied to their cases. 

We created a code implementing the methodology 
described above. We aim to make it public in the near future. 
To test the robustness of our implementation, 
we compared our results for the galaxy bias and 
$A_{\rm ISW}$ parameters with the ones in 
Ref.~\cite{Stolzner:2017ged} assuming the \lcdm\ model for which the best-fit cosmological parameters are fixed to the Planck 2015 results. 
We found that the galaxy bias factors are consistent with 
the values in Ref.~\cite{Stolzner:2017ged}. 
Furthermore, we confirmed that adopting either the values reported in Ref.~\cite{Stolzner:2017ged} or those computed with our code 
results in a negligible change for $A_{\rm ISW}$. 
To be concrete, we obtained the constraint 
$A_{\rm ISW} = 1.54 \pm 0.32$, which is within 
$0.1 \sigma$ of $A_{\rm ISW} = 1.51 \pm 0.30$ as 
reported in Ref.~\cite{Stolzner:2017ged}. 
Hence the codes used to compute these quantities 
are in agreement with each other.

Before concluding this subsection, we would like to stress that, 
instead of using the best-fit values in the analysis, one can 
alternatively resort to a full Markov Chain Monte Carlo (MCMC) 
approach that varies all of the parameters along with 
the galaxy bias factors and $A_{\rm ISW}$. 
In the \lcdm\ model, it has been 
found that both methods give the same 
conclusion \cite{Stolzner:2017ged}. 
In this work, we then adopt the best-fit approach for simplicity. 
We will leave a full MCMC analysis 
for a future work. 

\subsection{Best-fit values for cosmological and model parameters} \label{sec:bestfit}

For the purpose of the present work, it is sufficient to exploit only the best-fit values for the cosmologies we are interested in. 
They are obtained from minimization instead of full Markov Chain Monte Carlo (MCMC), as mentioned in the previous subsection. For this purpose, we run \eftcosmomc~\cite{Raveri:2014cka} and 
combine the following data sets: 

\begin{itemize}
\item \planck\ 2018 measurements of CMB temperature and polarization anisotropy \cite{Aghanim:2019ame}. 
In detail, we use the TT and EE power spectra, 
and TE cross-correlations over small angular scales ($\ell \in [30, 2508]$ for TT power spectrum, $\ell \in [30, 1996]$ for TE cross-correlation and EE power spectra). In all cases, we use the Pliklite likelihood, which marginalizes over foreground parameters.

\item Measurements of the BAO from BOSS DR12~\cite{Alam:2016hwk}, SDSS Main Galaxy Sample DR7~\cite{Ross2015}, and 6dFGS~\cite{Beutler2011}.

\item The 2018 Pantheon Supernova compilation~\cite{Scolnic_short/etal:2018}, which includes 1048 SNIa data in the redshift range $0.01< z <2.3$.

\end{itemize}

Hereafter, we refer to this combination of data as Baseline. Let us note that we do not include CMB lensing data by \planck\ because of the dependence of the lensing  potential power spectrum at multipole $\ell > 100$ on the parameters of the $\Lambda$CDM model \cite{Planck:2015mym}. For each minimization, we use the best-fit parameters found by an MCMC as a starting point for the minimization code. We use the built-in \eftcosmomc\ minimization code.
The best-fit parameters for the \lcdm, GGC, GCCG, and K-mouflage models 
are presented in Table~\ref{tab:BFParams}. 
In this table, $H_0=100\,h$~km\,sec$^{-1}$\,Mpc$^{-1}$ 
is today's Hubble parameter, $\Omega_{b,0}$ and 
$\Omega_{c,0}$ are today's density parameters of baryons 
and CDM respectively, $A_s$ and $n_s$ are the amplitude 
and scalar spectral index of primordial curvature 
perturbations, $\tau$ is the optical depth, 
and $S_8=\sigma_{8,0} \sqrt{\Omega_{{\rm m},0}/0.3}$, 
where $\sigma_{8,0}$ is today's amplitude of matter 
perturbations at the scale $8h^{-1}$~Mpc. 

For the K-mouflage minimization, we fixed three out of five model 
parameters to $m = 3$, $\gamma_A = 0.2$, and $\gamma_U = 1$ as they  are unconstrained by cosmological observables  \citep{Benevento:2018xcu}. We have also included the \lcdm\ best-fit values of cosmological parameters for reference. In all cases, we include massive neutrinos with fixed total mass of 
$\sum m_\nu=0.06\,\mathrm{eV}$.

Each MG model fits this combination of likelihoods slightly better than \lcdm\ as seen by the $\chi^2$ comparison in Table~\ref{tab:BFParams_chi2_breakdown}.
For the GGC and GCCG models, the reduction in $\chi^2$ 
with respect to \lcdm\ is $\Delta \chi^2=-7.6$ and $\Delta \chi^2=-5.1$, respectively. 
This comes primarily from a better fit to \planck\ CMB 
anisotropy data as shown in Table~\ref{tab:BFParams_chi2_breakdown} where it is clear that the actual improvement comes from the TTTEEE data for $\ell > 30$. This result has been previously discussed 
in Refs.~\cite{Peirone:2019aua,Frusciante:2019puu}. 
Since the inclusion of additional parameters might lead to a better 
fit to the data, only a proper model selection analysis based not only on the goodness of fit but also on the complexity of the model can 
actually inform if the model is preferred by data or not. 
This kind of analysis is out of the scope of the present work. 
However, a similar investigation has been performed for these two models when a proper MCMC analysis is adopted, 
and it has been found that for some combinations of data 
they are preferred over \lcdm~\cite{Peirone:2019aua,Frusciante:2019puu}. 

The best-fit K-mouflage model gives only a slight difference 
$\Delta \chi^2=-0.3$ relative to \lcdm, so it is essentially 
degenerate to the $\Lambda$CDM limit.
Therefore, it is unlikely that the combination of data sets 
used in our analysis will prefer K-mouflage over 
$\Lambda$CDM even by varying all five parameters.

\begin{table}[t!]
\centering
\begin{tabular}{|c|c|c|c|c|c|c|c|c|c|} 
\hline
Model & $H_0$ & $\Omega_{b,0} h^2$ & $\Omega_{c,0} h^2$ 
& $10^9 A_s$ & $n_s$ & $\tau$ & $S_8$ & 
Additional model parameters & $\chi^2$ \\
\hline
\lcdm\ & 67.69 & 0.02242 & 0.1193 & 2.099 & 0.9673 & 0.055 & 0.8234 & -- & 2044.8 \\
\hline
GGC & 68.24 & 0.02245 & 0.1196 & 2.108 & 0.9656 & 0.057 & 0.8502 & $x_1^0 = -1.20$, $x_3^0 = 0.38$ & 2037.2 \\
\hline
GCCG & 68.93 & 0.02240 & 0.1200 & 2.106 & 0.9656 & 0.056 & 0.8394 & $s = 0.182$, $q = 1.49$ & 2039.7\\
\hline
K-mouflage & 68.37 & 0.02241 & 0.1198 & 2.106 & 0.9670 & 0.056 & 0.8250 & $\alpha_U=0.505$,  $\epsilon_{2,0}=-6.9 \times 10^{-4}$ & 2044.5 \\
\hline
\end{tabular}
\caption{Best-fit cosmological parameters 
and corresponding $\chi^2$ values derived from likelihood minimization fits of the \lcdm, GGC, GCCG, and K-mouflage models, 
respectively (see Sec.~\ref{Sec:models}). 
For the K-mouflage case, we fix $\gamma_U=1$, $\gamma_A=0.2$, $m=3$ 
as these parameters are unconstrained by the likelihood. 
} \label{tab:BFParams}
\end{table}

\begin{table}[t!]
\centering
\begin{tabular}{|c|c|c|c|c|c|c|c|} 
\hline
Model & \planck\ 2018 TTTEEE $\ell > 30$ & \planck\ TT $\ell \leq 30$  & \planck\ LowE 
& SN & BAO & $\chi^2$ & $\Delta\chi^2$ \\
\hline
\lcdm\ & 584.9 & 23.0 & 396.1 & 1035.0 & 5.7 & 2044.8 & 0 \\
\hline
GGC & 578.0 & 22.3 & 396.6 & 1034.7 & 5.5 & 2037.2 & -7.6 \\
\hline
GCCG & 580.0 & 22.2 & 396.4 & 1035.0 & 6.132 & 2039.7 & -5.1 \\
\hline
K-mouflage & 584.7 & 23.2 & 396.4 & 1034.8 & 5.4 & 2044.5 & -0.3 \\
\hline
\end{tabular}
\caption{Breakdown in the $\chi^2$ by likelihood used for the best-fit \lcdm, GGC, GCCG, and 
K-mouflage models, 
respectively (see Sec.~\ref{Sec:models}). 
In the Table we define $\Delta \chi^2=
\chi^2_{\rm MG}-\chi^2_{\Lambda {\rm CDM}}$ 
at the difference between MG and \lcdm\, 
in the total $\chi^2$.} \label{tab:BFParams_chi2_breakdown}
\end{table}

\section{Results and Discussion}\label{Sec:Results}

We now confront theoretical predictions of 
the GGC, GCCG, and K-mouflage models with 
the measurements of ISW-galaxy cross-correlations.
For the purpose of understanding how the presence 
of additional model parameters to those in the 
$\Lambda$CDM model modifies the ISW-galaxy cross-correlation power spectra, 
we first show the theoretical values of $l (l+1)C_l^{\rm Tg}/(2\pi)$ 
for the same best-fit cosmological parameters as those 
in $\Lambda$CDM. 
Then, using the best-fit parameters derived 
using the Baseline likelihood for each model, 
we study whether the three classes of 
MG models can be compatible with 
the ISW-galaxy cross-correlation data. 
In the following, we will discuss each model 
in turn.

\subsection{Galileon Ghost Condensate (GGC)}
%

\begin{figure*}[t!]
\hspace*{-1cm}
\begin{center}
\includegraphics[width=7in, height=3in]{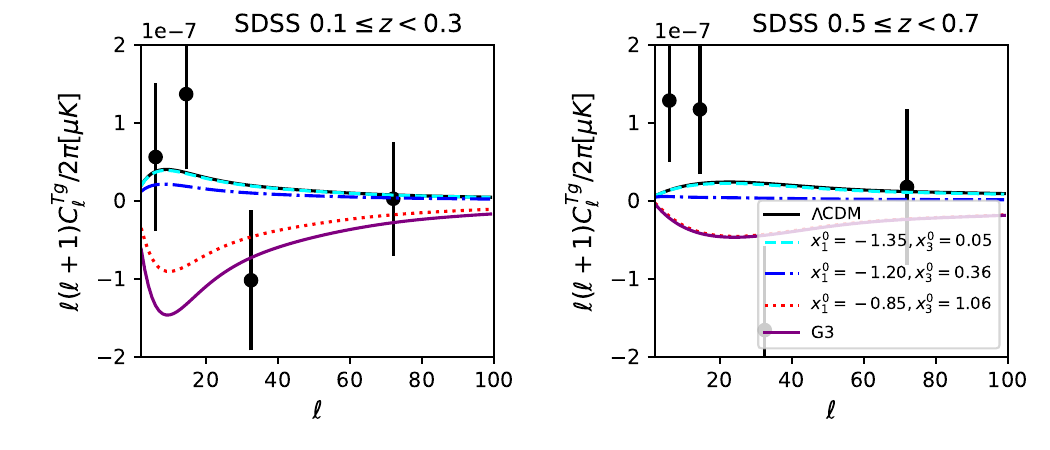}
\end{center}
\caption{In the GGC model, we plot the ISW-galaxy 
cross-correlation power spectra for two different 
redshift bins: (1) $0.1 \leq z<0.3$ (left) and 
(2) $0.5 \leq z<0.7$ (right).
We show $l(l+1)C_l^{\rm Tg}/(2\pi)$ for 
three different cases: 
(i) $x_1^0=-1.35$, $x_3^0=0.05$, 
(ii) $x_1^0=-1.20$, $x_3^0=0.36$, and 
(iii) $x_1^0=-0.85$, $x_3^0=1.06$, 
where $x_1^0$ and $x_3^0$ are 
today's values of $x_1$ and $x_3$ respectively.
The SDSS data are shown as black points 
with error bars. 
Besides the $\Lambda$CDM model, we also depict the 
cross-correlation power spectrum in the 
Cubic Galileon (G3) model.
In all cases, we fix the \lcdm\ model parameters to the \lcdm\ best-fit values presented in Table~\ref{tab:BFParams}. If the magnitude of $|x_1^0|$ increases relative to $x_3^0$, the ISW signal approaches the \lcdm\ limit. 
For $x_3^0$ exceeding the order of $|x_1^0|$, 
the ISW signal approaches the G3 limit. 
In this latter case, it is possible to generate a negative ISW-galaxy cross-correlation, which is disfavored when data from all redshift bins are included.}
\label{fig:GGC_theory}
\end{figure*}

\begin{table*}[ht!]
\centering
\begin{tabular}{|c|c|c|c|c|c|} 
\hline
Catalog & $z$ & \lcdm\ & GGC & GCCG & K-mouflage  \\
\hline
 & 0.1-0.3 & 1.16 & 1.12 & 1.12 & 1.15 \\
 & 0.3-0.4  & 1.04 & 1.00 & 1.01  & 1.03  \\
SDSS & 0.4-0.5 & 1.01 & 0.98 & 0.99 & 1.00  \\
 & 0.5-0.7 & 1.40 & 1.36 & 1.37 & 1.38  \\
 & 0.7-1.0 & 1.41 & 1.39 & 1.40 & 1.41  \\
\hline
 & 0.09-0.21 & 1.12 & 1.08 & 1.09 & 1.11 \\
WIxSC & 0.21-0.3 & 1.00 & 0.97 & 0.98 & 0.99 \\
 & 0.3-0.6 & 1.30 & 1.26 & 1.26 & 1.29  \\
\hline
 & 0.5-1 & 2.22 & 2.17 & 2.19 & 2.20 \\
QSO & 1-2 & 3.09 & 3.00 & 3.02 & 3.02 \\
 & 2-3 & 3.88 & 3.85 & 3.87 & 3.86 \\
\hline
 & 0-0.105 & 1.24 & 1.19 & 1.20 & 1.23  \\
2MPZ & 0.105-0.195 & 1.43 & 1.38 & 1.39 & 1.42  \\
 & 0.195-0.3 & 2.13 & 2.06 & 2.07 & 2.11 \\
\hline
NVSS & 0-5 & 2.22 & 2.15 & 2.16 & 2.20 \\
\hline
\end{tabular}
\caption{Best-fit galaxy bias factors $b$ 
for each redshift bin of five different catalogs. 
We consider four different cosmological models 
with fiducial parameters given 
in Table~\ref{tab:BFParams}.}
\label{tab:galaxy_bias_model}
\end{table*}

In the GGC model, the existence of the Galileon Lagrangian 
$3a_3 X \square \phi$ gives rise to a different cosmic 
growth history compared to $\Lambda$CDM. 
As we observe in Eq.~(\ref{muSigGa}), 
the variable $x_3$ leads to deviations of $\mu$ 
and $\Sigma$ from 1.
In Fig.~\ref{fig:GGC_theory}, we plot the ISW-galaxy 
cross-correlation power spectrum for several different 
sets of the GGC model parameters $x_1^0$ and 
$x_3^0$ (which are today's values of $x_1$ and $x_3$).
In the same figure, we also include the \lcdm\ and 
G3 models for comparison. 
To plot the power spectra for each model in Fig.~\ref{fig:GGC_theory}, 
we fix the \lcdm\ model 
parameters for each of the models depicted to the \lcdm\ best-fit values 
presented in Table~\ref{tab:BFParams}.

In Fig.~\ref{fig:GGC_theory}, we also show data from the cross-correlation of SDSS galaxy number counts with \planck\ temperature data in the redshift ranges $0.1 \leq z < 0.3$ and $0.5 \leq z < 0.7$. 
For the theoretical power spectra, we adopt the same selection 
function as used in the two redshift ranges of SDSS to elucidate the power spectrum for those particular redshift bins. 
Overall, the strength of  ISW signal is weaker in the higher redshift 
bin than in the lower redshift bin as the scalar field is subdominant 
to the background fluid at earlier times. 

\begin{figure*}[t!]
\hspace*{-1cm}
\begin{center}
\includegraphics[width=4.5in, height=2.7in]{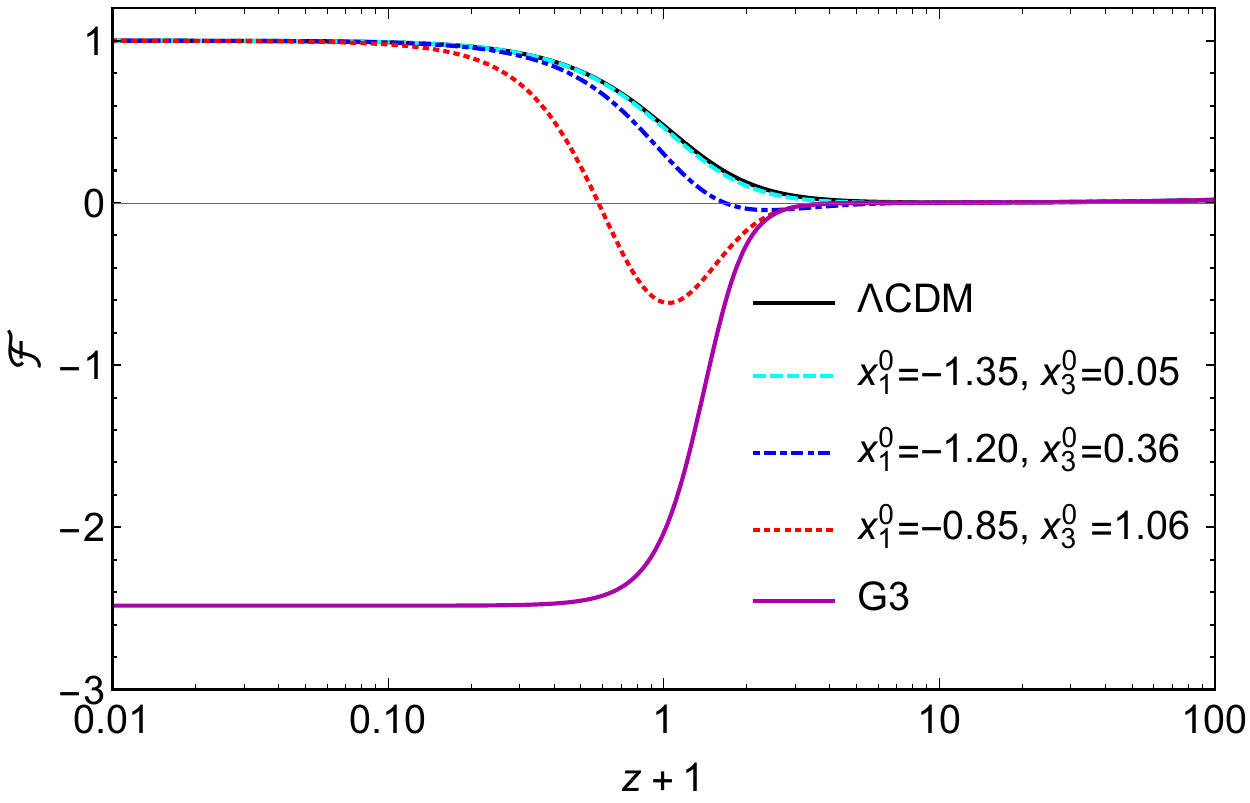}
\end{center}
\caption{Evolution of the quantity ${\cal F}$ versus 
$z+1$ for three different cases of the GGC model: 
(i) $x_1^0=-1.35$, $x_3^0=0.05$, 
(ii) $x_1^0=-1.20$, $x_3^0=0.36$, and 
(iii) $x_1^0=-0.85$, $x_3^0=1.06$. 
We include the plots of $\Lambda$CDM and G3 models 
as reference. As the ratio $x_3^0/|x_1^0|$ increases, ${\cal F}$ becomes negative in most of the low redshift range. A negative value of ${\cal F}$ is a necessary, though not sufficient condition, for a negative ISW-galaxy cross-correlation.}
\label{fig:F_GGC}
\end{figure*}

As we observe in Fig.~\ref{fig:GGC_theory}, irrespective of 
the redshift bins, larger values of $x_3^0$ work to suppress 
the ISW-galaxy cross-correlation power with respect to $\Lambda$CDM.
In the limit $x_3^0\rightarrow 0$, GGC mimics 
$\Lambda$CDM.
As $x_3^0$ exceeds the order of $|x_1^0|$, GGC approaches 
the G3 limit.
In Fig.~\ref{fig:F_GGC}, we plot the evolution of ${\cal F}$ 
defined by Eqn.~(\ref{calF}) for the same model parameters 
as those used in Fig.~\ref{fig:GGC_theory}.
{}From Eqn.~(\ref{CLTg2}), the negative ISW-galaxy 
cross-correlation is accompanied by negative values 
of ${\cal F}$, whose property can be confirmed by 
comparing Fig.~\ref{fig:GGC_theory} with Fig.~\ref{fig:F_GGC}. 
For increasing ratio $x_3^0/|x_1^0|$, ${\cal F}$ becomes 
negative in most of the low redshift ranges.
This leads to the suppressed ISW-galaxy cross-correlation power 
$l(l+1)C_l^{\rm Tg}/(2\pi)$, whose behavior is not favored from 
the SDSS data shown in Fig.~\ref{fig:GGC_theory}.

For the best-fit parameters presented in 
Table~\ref{tab:BFParams}, we compute the 
auto-correlation $\chi^2$ given by Eqn.~(\ref{auto_chi}).
Then, we find the best-fit galaxy bias parameters, 
which are given in Table~\ref{tab:galaxy_bias_model}. 
The obtained values of $b$ in GGC are similar to 
those in $\Lambda$CDM.
Using the best-fit parameters along with these galaxy 
bias values, we calculate the $\chi^2$ in 
Eqn.~\eqref{eq:chi2cc} for the ISW-galaxy 
cross-correlation power spectra. 
We consider the two cases: (1) $A_{\rm ISW}$ 
is fixed to 1, and (2) $A_{\rm ISW}$ is allowed to vary.
The $\chi^2$ values for cases (1) and (2) are denoted 
as $\chi_1^2$ and $\chi_{\rm A}^2$, respectively. 
The constrained values of $\chi_1^2$ and $\chi_{\rm A}^2$ are 
presented in Table~\ref{tab:BFchi2}. 
In $\Lambda$CDM the value of $A_{\rm ISW}$ derived by 
the ISW likelihood is slightly different from the one reported in Sec.~\ref{Sec:ISWsignificance}, 
because here we are using different 
fiducial \lcdm\ parameters. 

We find that the GGC best-fit model has a worse fit to the 
likelihood than \lcdm\ as quantified by an increase in both $\chi^2_1$ 
and $\chi^2_{\rm A}$. This increase of $\chi^2$ in GGC results from 
the reduction of power in the ISW signal. 
The suppressed ISW power manifests itself in the preference
for larger $A_{\rm ISW} = 2.71 \pm 0.94$ in 
comparison to the $\Lambda$CDM bound 
$A_{\rm ISW} = 1.61 \pm 0.33$.
The increases $\Delta \chi^2_1 = 15.7$ and $\Delta \chi_{\rm A}^2 = 15.3$ 
for the GGC best-fit relative to the \lcdm\ best-fit are greater than the 
reduction $\Delta \chi^2 = -7.6$ for the GGC best-fit relative to the \lcdm\ best-fit for the fit to the Baseline likelihood. 
This means that the GGC best-fit parameters shown in Table~\ref{tab:BFParams} are not compatible with  the ISW likelihood when combined with the Baseline data.

In Tables~\ref{tab:Chi2_breakdown} and \ref{tab:Chi2_breakdown_A}, 
we show the contributions to the $\chi^2$ arising from 
each redshift bin of the catalogs we are using. 
The largest contributions in both $\chi^2_1$ and $\chi^2_{\rm A}$ 
come from the higher redshift bins with NVSS alone,
contributing $\Delta \chi^2_1 = 5.92$ and $\Delta \chi^2_{\rm A} =6.66$, respectively. There are also increases of $\Delta \chi^2_1 > 1$ and 
$\Delta \chi^2_{\rm A}>1$ resulting from SDSS 
in the ranges $0.4 \leq z < 0.5$ and $0.5 \leq z < 0.7$ 
as well as QSO in the ranges $0.5 \leq z < 1$ and $1 \leq z < 2$. 
Excluding any one of the entire catalogs would result in
increases of $\chi^2_1$ and $\chi^2_{\rm A}$. 
This increase is sufficient to offset the reduction of $\chi^2$ shown in Table \ref{tab:BFParams_chi2_breakdown}. Therefore, 
the fact that the GGC best-fit model is incompatible with the ISW likelihood 
in comparison to the \lcdm\ best-fit model
is not associated with 
a systematic error in one of the data sets 
or redshift bins. 
Additionally, even when the phenomenological 
rescaling of $A_{\rm ISW}$ is used, 
the $\chi^2$ in GGC is significantly larger than 
that in $\Lambda$CDM. This suggests that 
the worse fit 
relative to \lcdm\ is not only the result of a reduction of the ISW power, but also of a change in the shape of the power spectrum. In particular we find that, for high redshift measurements of QSO and NVSS, the best-fit GGC model has a negative ISW-galaxy cross-correlation.

\begin{table*}[t!]
\centering
\begin{tabular}{|c||c|c||c|c|} 
\hline
Model &$A_{\rm ISW}$ &$\chi^2_{1}$ &$A_{\rm ISW}$ &  $\chi^2_{\rm A}$ \\
\hline
\hline
\lcdm\ &1 & 47.1& $1.61 \pm 0.33$  & 43.4\\
\hline
GGC &1 & 62.8& $2.71 \pm 0.94$  & 58.7  \\
\hline
GCCG &1& 58.6& $4.22 \pm 0.90$  & 44.9 \\
\hline
K-mouflage &1& 46.9 & $1.59 \pm 0.33$ & 43.4 \\
\hline
\end{tabular}
\caption{For four different cosmological models, 
we show $\chi^2_1$ for $A_{\rm ISW}=1$ and $\chi^2_{\rm A}$ 
when $A_{\rm ISW}$ is allowed to vary.
They are derived by using best-fit cosmological 
parameters extracted from the likelihood minimization 
(see Sec.~\ref{sec:bestfit}).} 
\label{tab:BFchi2}
\end{table*}

\begin{table*}[ht!]
\centering
\begin{tabular}{|c|c|c|c|c|c|} 
\hline
Catalog & $z$ & $\chi^2_{1,\Lambda {\rm CDM}}$ & 
$\Delta \chi^2_{1,{\rm GGC}}$ & 
$\Delta \chi^2_{1,{\rm GCCG}}$ & 
$\Delta \chi^2_{1,{\rm KM}}$ \\
\hline
 & 0.1-0.3 & 2.85 & 0.08 & 0.13 & 0.00\\
 & 0.3-0.4  & 1.59 & 0.85 & 0.80 & -0.01 \\
SDSS & 0.4-0.5 & 2.50 & 1.64 & 1.42 & -0.02 \\
 & 0.5-0.7 & 5.95 & 2.51 & 1.98 & -0.04 \\
 & 0.7-1.0 & 6.42 & 0.09 & 0.16 & 0.00 \\
\hline
 & 0.09-0.21 & 2.34 & -0.18 & -0.20 & 0.01\\
WIxSC & 0.21-0.3 & 3.10 & 0.41 & 0.49 & -0.01\\
 & 0.3-0.6 & 1.98 & 0.06 & 0.05 & 0.00 \\
\hline
 & 0.5-1 & 5.38 & 1.72 & 1.17 & -0.02\\
QSO & 1-2 & 1.31 & 1.80 & 0.78 & -0.01\\
 & 2-3 & 3.67 & 0.46 & 0.16 & 0.00\\
\hline
 & 0-0.105 & 2.04 & 0.05 & 0.07 & 0.00 \\
2MPZ & 0.105-0.195 & 0.67 & 0.05 & 0.09 & 0.00 \\
 & 0.195-0.3 & 0.69 & 0.24 & 0.29 & 0.00\\
\hline
NVSS & 0-5 & 6.56 & 5.92 & 4.14 & -0.06\\
\hline
\end{tabular}
\caption{The differences of $\chi_1^2$ between 
three MG models and $\Lambda$CDM, i.e., 
$\Delta \chi^2_{1,x} = \chi^2_{1,x} - 
\chi^2_{1,\Lambda {\rm CDM}}$, 
where $x$ denotes either GGC, GCCG, 
K-mouflage (KM) models.
The $\chi_{\rm 1}^2$ values in $\Lambda$CDM are 
also presented.
In all cases, $A_{\rm ISW} = 1$. 
We show the values of $\Delta \chi^2_{1,x}$ for
each redshift bin of five different catalogs of 
the ISW-galaxy cross-correlation likelihood.}
\label{tab:Chi2_breakdown}
\end{table*}

\begin{table*}[ht!]
\centering
\begin{tabular}{|c|c|c|c|c|c|} 
\hline
Catalog & $z$ & $\chi^2_{{\rm A},\Lambda {\rm CDM}}$ & 
$\Delta \chi^2_{{\rm A},{\rm GGC}}$ & $\Delta 
\chi^2_{{\rm A},{\rm GCCG}}$ & $\Delta 
\chi^2_{{\rm A},{\rm KM}}$ \\
\hline
 & 0.1-0.3 & 3.22 & -0.28 & 0.09 & 0.00\\
 & 0.3-0.4  & 1.33 & 0.39 & 0.02 & 0.00 \\
SDSS & 0.4-0.5 & 1.72 & 1.36 & 0.18 & 0.00 \\
 & 0.5-0.7 & 4.86 & 2.54 & 0.32 & -0.01 \\
 & 0.7-1.0 & 6.41 & -0.38 & -0.25 & 0.00 \\
\hline
 & 0.09-0.21 & 2.78 & -0.13 & 0.20 & 0.00 \\
WIxSC & 0.21-0.3 & 3.10 & -0.14 & -0.04 & 0.00 \\
 & 0.3-0.6 & 2.34 & -0.35 & -0.07 & 0.00 \\
\hline
 & 0.5-1 & 4.86 & 1.90 & 0.10 & 0.00 \\
QSO & 1-2 & 1.05 & 3.00 & -0.06 & 0.00\\
 & 2-3 & 3.59 & 0.89 & 0.05 & 0.00\\
\hline
 & 0-0.105 & 1.99 & -0.01 & -0.01 & 0.00 \\
2MPZ & 0.105-0.195 & 0.76 & -0.02 & 0.09 & 0.00 \\
 & 0.195-0.3 & 0.75 & -0.06 & 0.05 & 0.00\\
\hline
NVSS & 0-5 & 4.62 & 6.66 & 0.88 & 0.00 \\
\hline
\end{tabular}
\caption{
The differences of $\chi_{\rm A}^2$ between 
three MG models and $\Lambda$CDM, i.e., 
$\Delta \chi^2_{{\rm A},x} = 
\chi^2_{{\rm A},x} - 
\chi^2_{{\rm A},\Lambda {\rm CDM}}$, 
where $x$ denotes either GGC, GCCG, 
K-mouflage (KM) models. 
The $\chi_{\rm A}^2$ values in $\Lambda$CDM are 
also shown.
In all cases, $A_{\rm ISW}$'s are fixed to 
the best-fit values shown in Table \ref{tab:BFchi2}.
We present the values of $\Delta \chi^2_{{\rm A},x}$ 
for each redshift bin of five different catalogs of 
the ISW-galaxy cross-correlation likelihood.}
\label{tab:Chi2_breakdown_A}
\end{table*}

\subsection{Generalized Cubic Covariant Galileon (GCCG)}

\begin{figure*}[t!]
\hspace*{-1cm}
\includegraphics[width=7in, height=3in]{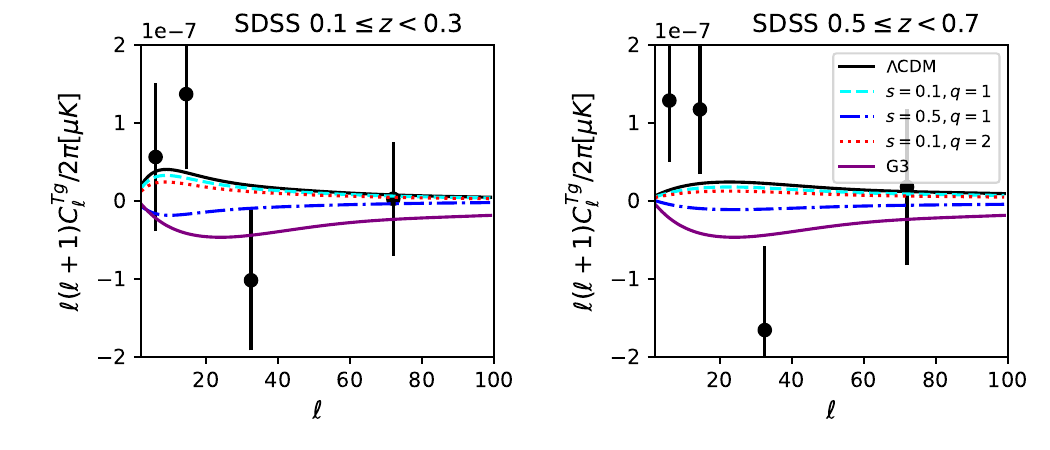}
\caption{The ISW-galaxy cross-correlation power spectra 
for the GCCG model for two redshift bins. 
The data points are the same as those in Fig.~\ref{fig:GGC_theory}.
Each case corresponds to the model parameters 
(i) $s=0.1$, $q=1$, (ii) $s=0.5$, $q=1$, and 
(iii) $s=0.1$, $q=2$. We also plot the 
cross-correlation power spectra in $\Lambda$CDM 
and G3 models.
In all cases, we fix the parameters present in 
the $\Lambda$CDM model to the \lcdm\ best-fit values 
given in Table \ref{tab:BFParams}. 
For increasing values of $s$ or $q$ the evolution of 
perturbations exhibits a larger deviation from 
that in $\Lambda$CDM, in which case it is possible to 
generate negative ISW-galaxy cross-correlations.
This negative cross-correlation is disfavored when 
data from all redshift bins are included. }
\label{fig:GCCG_theory}
\end{figure*}

As we discussed in Sec.~\ref{GCCG_model}, 
the key quantities of the GCCG model affecting 
the cosmic expansion and growth histories are 
the two parameters $q$ and $s=p/q$. 
At the background level, the deviation of $w_{\rm DE}$ from 
$-1$ is quantified by the parameter $s$. 
For the tracker solution, the deviations of $\mu$ and 
$\Sigma$ from 1 are proportional to $4s q^2 r_2$, 
see Eqn.~(\ref{musitra}). 
This means that, for increasing values of $s$ or $q$, 
the evolution of perturbations is generally subject to 
modifications in comparison to $\Lambda$CDM.

In Fig.~\ref{fig:GCCG_theory}, we depict the ISW-galaxy 
cross-correlation power spectra for the GCCG model for 
several different values of $s$ and $q$. 
We also include the plots for 
the \lcdm\ and G3 models, together with data from 
the SDSS catalog. For the theory power spectra, we use the same selection function as used by the SDSS redshift bins.
In this figure, we choose cosmological parameters present in
both $\Lambda$CDM and MG models to be the 
\lcdm\ best-fit values given in Table~\ref{tab:BFParams}. 
Depending on the values of $s$ and $q$, 
it is possible to realize both positive and negative 
cross-correlations. 
As we see in Fig.~\ref{fig:GCCG_theory}, increasing 
the values of either $s$ or $q$ 
works to suppress the ISW-galaxy cross-correlation 
power relative to $\Lambda$CDM. 
In particular, the cross-correlation becomes negative
for large products $sq$.

As in the GGC model, we evaluate the best-fit values of 
bias factors $b$ according to Eqn.~(\ref{auto_chi}) 
by adopting the best-fit model and cosmological parameters given in Table~\ref{tab:BFParams}. 
The best-fit values of $b$ in the GCCG, which are 
presented in Table~\ref{tab:galaxy_bias_model}, 
are similar to those in the $\Lambda$CDM case.
We also compute the $\chi^2$ of the ISW-galaxy cross-correlation power spectra given by Eqn.~\eqref{eq:chi2cc}. 
In Table \ref{tab:BFchi2} we show the values of $\chi_1^2$ and $\chi_{\rm A}^2$, which are derived by 
setting $A_{\rm ISW}=1$ and by varying $A_{\rm ISW}$ 
respectively.
The $\chi^2$ for the Baseline likelihood alone exhibits 
a better fit to data relative to \lcdm\ ($\Delta \chi^2 = -5.1$).
Taking into account the ISW-galaxy cross-correlation data; 
however, the $\chi^2_1$ in GCCG is larger than the \lcdm\ value, with difference $\Delta \chi^2_1 = 11.5$. This suggests that GCCG best-fit model given in Table~\ref{tab:BFParams} is not compatible with
the ISW-galaxy likelihood when combined with the Baseline data.  

Allowing $A_{\rm ISW}$ to vary, we find that 
the difference of $\Delta \chi^2_{\rm A}$ between 
the best-fit GCCG and the best-fit \lcdm\ models is $\Delta \chi^2_{\rm A} 
= 1.5$, which is smaller than $\Delta \chi^2_1 = 11.5$ 
obtained for $A_{\rm ISW}=1$.
The constraint on $A_{\rm ISW}$ is 
$A_{\rm ISW} = 4.22 \pm 0.90$, so the best-fit GCCG 
model predicts a smaller ISW  
power compared to the \lcdm\ model.
Since the difference $\Delta \chi^2_{\rm A}$ between the best-fit GCCG and best-fit \lcdm\ models is closer to zero when $A_{\rm ISW}$ is allowed to vary, it follows that the 
suppression of ISW power for the best-fit GCCG is the primary reason for the incompatibility with 
the ISW-galaxy data relative to \lcdm. 

In Tables~\ref{tab:Chi2_breakdown} and \ref{tab:Chi2_breakdown_A}, 
we show the relative differences of $\chi^2_1$ as well as $\chi_{\rm A}^2$ 
between three MG models and $\Lambda$CDM 
for each redshift bin of the five catalogs. 
As in the best-fit GGC model, the preference for larger values of $\chi^2_1$ in the best-fit GCCG arises from the high redshift bins with NVSS alone contributing 
$\Delta\chi^2_1 = 4.14$. There are also increases in $\Delta \chi^2_1 > 1$ resulting from SDSS in the ranges $0.4 \leq z < 0.5$ and $0.5 \leq z < 0.7$ as well as QSO in the region $0.5 \leq z < 1$. 
Removing any one of the entire photometric galaxy catalogs still results in an increase of $\chi^2_1$ greater than the improvement of the best-fit GCCG in the $\chi^2$ fit to the Baseline likelihood relative to the best-fit $\Lambda$CDM case. 
This suggests that the incompatibility of GCCG best-fit to the 
ISW-galaxy cross-correlation likelihood relative to \lcdm\ best-fit is not attributed to a systematic error in any one of the data sets or redshift bins. 

\subsection{K-mouflage}

\begin{figure*}[t!]
\hspace*{-1cm}
\includegraphics[width=7in, height=3in]{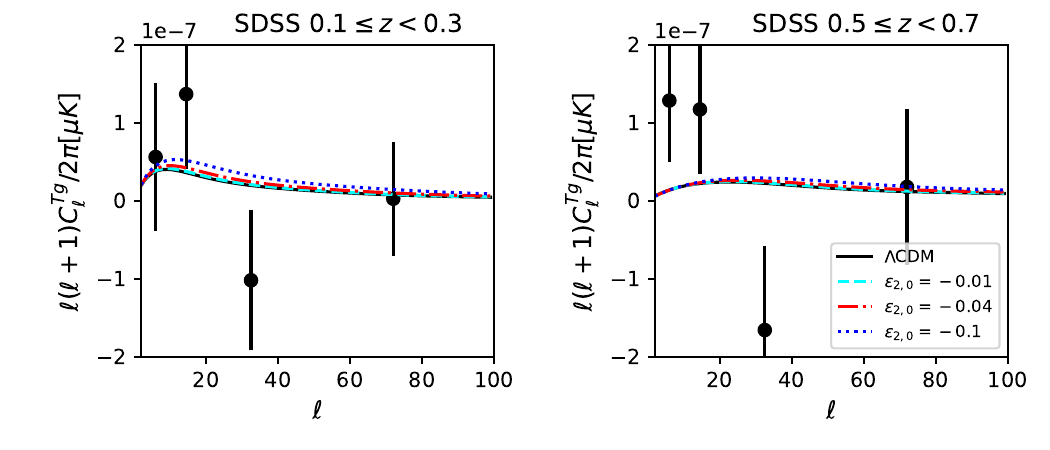}
\caption{The ISW-galaxy cross-correlation power spectra for two redshift bins 
in the K-mouflage model. The data points are the same as those 
in Fig.~\ref{fig:GGC_theory}. 
Each case corresponds to the model parameters 
(i) $\epsilon_{2,0}=-0.01$, 
(ii) $\epsilon_{2,0}=-0.04$, and 
(iii) $\epsilon_{2,0}=-0.1$, together with the
plot of the $\Lambda$CDM model (which corresponds to 
the limit $\epsilon_{2,0} \to 0$). 
In all cases, the parameters present in $\Lambda$CDM 
are fixed to the \lcdm\ best-fit values shown in Table~\ref{tab:BFParams}. 
For decreasing $|\epsilon_{2,0}|$, the ISW-galaxy cross-correlation 
power spectra are subject to the mild enhancement in comparison 
to those in $\Lambda$CDM.
At higher redshifts, the differences between 
K-mouflage and \lcdm\ models tend to be smaller.}
\label{fig:Kmouflage_theory}
\end{figure*}

As we discussed in Sec.~\ref{K-mouflage_model}, a key quantity that determines  the cosmic expansion and growth histories in the K-mouflage model 
is the parameter $\epsilon_{2,0}$.
In Fig.~\ref{fig:Kmouflage_theory} we show the theoretical predictions of the ISW-galaxy cross-correlation power spectra for several choices of $\epsilon_{2,0}$, together with the SDSS data in two different 
redshift bins. We fix parameters present in the \lcdm\ model, to the \lcdm\ best-fit values in Table \ref{tab:BFParams}. We use the same selection function for the theory power as used by the SDSS data in the specific redshift bin. We observe that larger negative values of $\epsilon_{2,0}$ 
result in an increase of the ISW-galaxy cross-correlation power 
spectra compared to those in $\Lambda$CDM 
(which corresponds to the limit $\epsilon_{2,0} \to 0$). 
This is associated with the fact that the quantity $\Sigma$ decreases 
in time for $\epsilon_2<0$, so the ISW-galaxy 
cross-correlation is always positive. 
Moreover, the quantities $\mu=(1+\epsilon_1)A^2$ and $\Sigma=A^2$ are 
larger than 1 in the past for the normalization $A(\phi_0)=1$ today. 
Since the variation of $A(\phi)$ occurs mostly at late times, 
the deviations of $l(l+1)C_l^{\rm Tg}/(2\pi)$ from the $\Lambda$CDM case 
in the redshift range $0.1 \le z<0.3$ are more significant that those 
in the range $0.5 \le z<0.7$.

The maximum likelihood Baseline likelihood analysis shows that the best-fit value of 
$\epsilon_{2,0}$ is $-6.9 \times 10^{-4}$ (see Table \ref{tab:BFParams}), 
which is relatively close to zero. 
Hence only minimal deviations from the \lcdm\ model are allowed from 
joint constraints with the CMB, BAO, and SNIa data for both 
the background and perturbations.
This phenomenology improves the fit to the ISW cross-correlation 
between CMB temperature anisotropy and galaxy number counts. 
For $A_{\rm ISW}=1$, as shown in Table~\ref{tab:BFchi2}, 
the $\chi^2$ value is improved only by $\Delta \chi_1^2=-0.2$ 
compared to $\Lambda$CDM. 
If $A_{\rm ISW}$ is allowed to vary, the $\chi^2_{\rm A}$ 
values for K-mouflage and $\Lambda$CDM are exactly the same for their best-fit parameters. 
This suggests that the only difference between the two model predictions 
comes from a slight preference for more ISW power in the K-mouflage case. 
In terms of $A_{\rm ISW}$ the K-mouflage model predicts 
$A_{\rm ISW}=1.59\pm 0.33$, while the $\Lambda$CDM value 
$A_{\rm ISW}=1.61\pm 0.33$ is slightly larger.

In the future, we plan to use more updated catalogs such as the DESI data \cite{Hang/etal2021}, especially in light of the minor differences in the measured values of $A_{\rm ISW}$ between DESI and our collection of catalogs for the \lcdm\ scenario. Indeed, according to the results we obtained for the $\Lambda$CDM model 
with the Planck 2018 best-fit values, we find $A_{\rm ISW} > 1$ 
at $1.85 \sigma$. This implies that \lcdm\ is predicting less power for the ISW-galaxy cross-correlation power spectrum than is being measured.
In Ref.~\cite{Hang/etal2021}, this preference for $A_{\rm ISW}>1$ is not found in the cross-correlation between the DESI Legacy Survey and Planck 2018 data. Constraints from DESI would be an interesting test for MG models. 
%

\section{Conclusion}\label{Sec:Conclusion}

In this paper, we studied the ISW-galaxy cross-correlation to probe 
three MG models of dark energy, i.e., GGC, GCCG, and K-mouflage models.
For this purpose, we have used a tomographic analysis of the ISW signal 
based on photometric measurements of the redshift of galaxies 
following the prescription in Ref.~\cite{Stolzner:2017ged}. 
The previous analysis assumed the standard flat $\Lambda$CDM cosmological model, so we extended it to include MG models in the framework 
of Horndeski theories. The GGC and GCCG models contain the cubic-order 
derivative interaction $G_3(X) \square \phi$, which can induce 
negative ISW-galaxy cross-correlations. On the other hand, 
the K-mouflage model gives rise to only the positive cross-correlation. 
These properties can be used to distinguish between different MG models 
from the ISW-galaxy cross-correlation power spectrum.

In our analysis, we first found the best-fit parameters for each of the cosmological models derived by joint constraints with the 
CMB, BAO and SNIa data. They are given in Table~\ref{tab:BFParams} 
for the GGC, GCCG, and K-mouflage models besides $\Lambda$CDM.
Then, these best-fit parameters were used as a fiducial model to 
compute the $\chi^2$ values in a fit to the ISW-galaxy 
cross-correlation and the collection of photometric redshift survey. 
This analysis was performed for two cases; one where we included a phenomenological rescaling of the theoretical cross-correlation 
curve via a parameter $A_{\rm ISW}$, and the other where 
$A_{\rm ISW}$ is fixed to one. 
In the case where $A_{\rm ISW}$ is allowed to vary, 
we summarize the best-fit values of $A_{\rm ISW}$ 
for each model in Table~\ref{tab:BFchi2}.

When $A_{\rm ISW}$ is kept fixed to 1, the GGC and 
GCCG best-fit models are incompatible with the ISW-galaxy cross-correlation data relative to the \lcdm\ best-fit model. 
This is mostly attributed to the fact that 
the best-fit values 
for the GGC and GCCG models prefer a suppressed ISW signal compared to \lcdm, 
which reduces the power in the ISW-galaxy cross-correlation. 
As we observe in Figs.~\ref{fig:GGC_theory} 
and \ref{fig:GCCG_theory}, these models can give rise to 
negative ISW-galaxy cross-correlations when the quantity $\Sigma$ 
grows rapidly to reach the regime ${\cal F}<0$. 
Indeed, this increase of $\Sigma$ induced by the cubic 
coupling $G_3(X) \square \phi$ leads to the suppressed ISW-galaxy cross-correlation power spectrum in comparison to $\Lambda$CDM, 
whose property tends to be incompatible with the observational data.

For the GCCG best-fit model, the suppressed ISW signal can be offset by a larger value of 
$A_{\rm ISW}$ than in $\Lambda$CDM. For the GGC best-fit model, allowing 
$A_{\rm ISW}$ to vary for the best-fit parameters derived by Baseline 
likelihood is not sufficient to bring it into comparable agreement with the ISW-galaxy cross-correlation data as the \lcdm\ best-fit case.
These results show that, while the suppression of ISW power 
in GCCG is the main reason for the tension with data, 
the modification of ISW-galaxy cross-correlation power spectrum 
itself in GGC largely contributes to the incompatibility with data.
The K-mouflage model has best-fit parameters whose deviations from 
$\Lambda$CDM are small, in addition to the fact that the ISW power 
is always positively correlated with galaxy fluctuations. 
We find that K-mouflage cannot be distinguished from $\Lambda$CDM 
even by adding the ISW-galaxy cross-correlation data.

In the future, we plan to perform a further exploration with 
a proper MCMC analysis to constrain the whole parameter space. 
This will allow us to perform a model selection analysis and 
see whether the GCG and GCCG models are ruled out or not 
by ISW-galaxy cross-correlation data. 
Furthermore, as more and more photometric measurements 
of redshift of galaxies are gathered, the errors from 
the ISW-galaxy cross-correlation will be reduced.
With such future high-precision measurements, it will be 
possible to scrutinize the possible large-distance modifications of 
gravity.

\acknowledgments

JK and GB wish to thank Charles Bennett, Graeme Addison, and Janet Weiland for many stimulating discussions during the analysis and write up for this work. Additionally, we wish to thank Simone Peirone for collaboration in the early stages of the work.
For JK and GB, this works was performed, in part, for the Jet Propulsion Laboratory, California Institute of Technology, sponsored by the United States Government under the Prime Contract 80NM00018D0004 between Caltech and NASA under subcontract numbers 1643587 and 1658514. This research project was conducted using computational resources at the Maryland Advanced Research Computing Center (MARCC). We acknowledge the use of the Legacy Archive for Microwave Background Data Analysis (LAMBDA), part of the High Energy Astrophysics Science Archive Center (HEASARC). HEASARC/LAMBDA is a service of the Astrophysics Science Division at the NASA Goddard Space Flight Center. NF is supported by Funda\c{c}\~{a}o para a Ci\^{e}ncia e a Tecnologia (FCT) through the research grants UIDB/04434/2020, UIDP/04434/2020, PTDC/FIS-OUT/29048/2017, CERN/FIS-PAR/0037/2019 and the personal research FCT grant ``CosmoTests -- Cosmological tests of gravity theories beyond General Relativity" with ref.~number CEECIND/00017/2018. The work of ADF \ was supported by Japan Society for the Promotion of Science~(JSPS) Grants-in-Aid for Scientific Research No.\ 20K03969. 
ST is supported by the Grant-in-Aid for Scientific Research Fund 
of the JSPS No.\,19K03854.

\bibliography{references}

\end{document}